\documentclass[aps,prd,twocolumn,superscriptaddress,showpacs,bibnotes]{revtex4}

\usepackage{graphicx}

\begin{document}

\title{First search for atmospheric and extraterrestrial neutrino-induced cascades with the IceCube detector}

\affiliation{III. Physikalisches Institut, RWTH Aachen University, D-52056 Aachen, Germany}
\affiliation{Dept.~of Physics and Astronomy, University of Alabama, Tuscaloosa, AL 35487, USA}
\affiliation{Dept.~of Physics and Astronomy, University of Alaska Anchorage, 3211 Providence Dr., Anchorage, AK 99508, USA}
\affiliation{CTSPS, Clark-Atlanta University, Atlanta, GA 30314, USA}
\affiliation{School of Physics and Center for Relativistic Astrophysics, Georgia Institute of Technology, Atlanta, GA 30332, USA}
\affiliation{Dept.~of Physics, Southern University, Baton Rouge, LA 70813, USA}
\affiliation{Dept.~of Physics, University of California, Berkeley, CA 94720, USA}
\affiliation{Lawrence Berkeley National Laboratory, Berkeley, CA 94720, USA}
\affiliation{Institut f\"ur Physik, Humboldt-Universit\"at zu Berlin, D-12489 Berlin, Germany}
\affiliation{Fakult\"at f\"ur Physik \& Astronomie, Ruhr-Universit\"at Bochum, D-44780 Bochum, Germany}
\affiliation{Physikalisches Institut, Universit\"at Bonn, Nussallee 12, D-53115 Bonn, Germany}
\affiliation{Dept.~of Physics, University of the West Indies, Cave Hill Campus, Bridgetown BB11000, Barbados}
\affiliation{Universit\'e Libre de Bruxelles, Science Faculty CP230, B-1050 Brussels, Belgium}
\affiliation{Vrije Universiteit Brussel, Dienst ELEM, B-1050 Brussels, Belgium}
\affiliation{Dept.~of Physics, Chiba University, Chiba 263-8522, Japan}
\affiliation{Dept.~of Physics and Astronomy, University of Canterbury, Private Bag 4800, Christchurch, New Zealand}
\affiliation{Dept.~of Physics, University of Maryland, College Park, MD 20742, USA}
\affiliation{Dept.~of Physics and Center for Cosmology and Astro-Particle Physics, Ohio State University, Columbus, OH 43210, USA}
\affiliation{Dept.~of Astronomy, Ohio State University, Columbus, OH 43210, USA}
\affiliation{Dept.~of Physics, TU Dortmund University, D-44221 Dortmund, Germany}
\affiliation{Dept.~of Physics, University of Alberta, Edmonton, Alberta, Canada T6G 2G7}
\affiliation{Dept.~of Subatomic and Radiation Physics, University of Gent, B-9000 Gent, Belgium}
\affiliation{Max-Planck-Institut f\"ur Kernphysik, D-69177 Heidelberg, Germany}
\affiliation{Dept.~of Physics and Astronomy, University of California, Irvine, CA 92697, USA}
\affiliation{Laboratory for High Energy Physics, \'Ecole Polytechnique F\'ed\'erale, CH-1015 Lausanne, Switzerland}
\affiliation{Dept.~of Physics and Astronomy, University of Kansas, Lawrence, KS 66045, USA}
\affiliation{Dept.~of Astronomy, University of Wisconsin, Madison, WI 53706, USA}
\affiliation{Dept.~of Physics, University of Wisconsin, Madison, WI 53706, USA}
\affiliation{Institute of Physics, University of Mainz, Staudinger Weg 7, D-55099 Mainz, Germany}
\affiliation{Universit\'e de Mons, 7000 Mons, Belgium}
\affiliation{Bartol Research Institute and Department of Physics and Astronomy, University of Delaware, Newark, DE 19716, USA}
\affiliation{Dept.~of Physics, University of Oxford, 1 Keble Road, Oxford OX1 3NP, UK}
\affiliation{Dept.~of Physics, University of Wisconsin, River Falls, WI 54022, USA}
\affiliation{Oskar Klein Centre and Dept.~of Physics, Stockholm University, SE-10691 Stockholm, Sweden}
\affiliation{Dept.~of Astronomy and Astrophysics, Pennsylvania State University, University Park, PA 16802, USA}
\affiliation{Dept.~of Physics, Pennsylvania State University, University Park, PA 16802, USA}
\affiliation{Dept.~of Physics and Astronomy, Uppsala University, Box 516, S-75120 Uppsala, Sweden}
\affiliation{Dept.~of Physics, University of Wuppertal, D-42119 Wuppertal, Germany}
\affiliation{DESY, D-15735 Zeuthen, Germany}

\author{R.~Abbasi}
\affiliation{Dept.~of Physics, University of Wisconsin, Madison, WI 53706, USA}
\author{Y.~Abdou}
\affiliation{Dept.~of Subatomic and Radiation Physics, University of Gent, B-9000 Gent, Belgium}
\author{T.~Abu-Zayyad}
\affiliation{Dept.~of Physics, University of Wisconsin, River Falls, WI 54022, USA}
\author{J.~Adams}
\affiliation{Dept.~of Physics and Astronomy, University of Canterbury, Private Bag 4800, Christchurch, New Zealand}
\author{J.~A.~Aguilar}
\affiliation{Dept.~of Physics, University of Wisconsin, Madison, WI 53706, USA}
\author{M.~Ahlers}
\affiliation{Dept.~of Physics, University of Oxford, 1 Keble Road, Oxford OX1 3NP, UK}
\author{K.~Andeen}
\affiliation{Dept.~of Physics, University of Wisconsin, Madison, WI 53706, USA}
\author{J.~Auffenberg}
\affiliation{Dept.~of Physics, University of Wuppertal, D-42119 Wuppertal, Germany}
\author{X.~Bai}
\affiliation{Bartol Research Institute and Department of Physics and Astronomy, University of Delaware, Newark, DE 19716, USA}
\author{M.~Baker}
\affiliation{Dept.~of Physics, University of Wisconsin, Madison, WI 53706, USA}
% ...

\collaboration{IceCube Collaboration}\noaffiliation 

\date{\today}

\begin{abstract}
We report on the first search for atmospheric and for diffuse astrophysical neutrino-induced showers (cascades) in the IceCube detector using 257 days
of data  collected in the year 2007-2008 with 22 strings active.
A total of 14 events with energies above 16\,TeV remained after event selections in the diffuse analysis, with an expected 
total background contribution of $8.3\pm 3.6$.
At 90\% confidence we set an upper limit of $E^2\Phi_{90\%CL}<3.6\times10^{-7}\,\mathrm{GeV \cdot cm^{-2} \cdot s^{-1}\cdot sr^{-1}}$ on the diffuse flux of neutrinos of all flavors 
in the energy range between $24$\,TeV and $6.6$\,PeV assuming that $\Phi \propto E^{-2}$ and that 
the flavor composition of the $\nu_e : \nu_\mu : \nu_\tau$ flux is $1 : 1 : 1$ at the Earth.
The atmospheric neutrino analysis was optimized for lower energies.  A total of 12 events were observed with energies above 5 TeV.
The observed number of events is consistent with the expected background, within the uncertainties.

\end{abstract}

\maketitle


\section{Introduction}
The origin of high-energy cosmic-rays is an area of active research in astrophysics.
The sites where cosmic rays are accelerated are expected to produce high energy neutrinos.
Many types of objects, ranging from supernovae and gamma-ray bursters to active galactic nuclei~\cite{theory}, have been proposed as point sources of high energy neutrinos 
and many searches for such sources have been made~\cite{ICSearches},  
yielding results consistent with background only assumptions.
If there are many point sources, each with an unobservably low flux, then the aggregate flux may still be observable as a diffuse flux.

Diffuse searches rely on the energy spectrum of the detected events to separate an extraterrestrial signal from atmospheric neutrinos produced in the interaction of cosmic rays with atomic nuclei in the Earth's atmosphere.
Low energy (below $\sim10$\,GeV)  atmospheric muon and electron neutrinos have been observed in underground detectors~\cite{lowenu}. 
At higher energies, from 100\,GeV to 400\,TeV, 
neutrino telescopes have measured the spectrum of atmospheric $\nu_\mu$~\cite{IceCube2010}.
 In this energy range, the flux of $\nu_e$ is expected to be lower by about a factor of 20~\cite{beacom}  and has not  been observed.

The main component of the atmospheric neutrino spectrum is produced by the decays of $\pi^\pm$ and $K^\pm$. 
Asymptotically it can be parametrized by $dN_\nu/dE_\nu \propto E_\nu^{-3.7}$, where $E_\nu$ is the neutrino energy~\cite{ICCRspectrum}.
Decays of hadrons containing charm and bottom quarks form an additional component that is expected to be close to the primary cosmic-ray spectrum, 
$dN_\nu/dE_\nu \propto E_\nu^{-2.7}$~\cite{naumov,enberg,martin}, and produces nearly equal numbers of $\nu_\mu$ and $\nu_e$.
These prompt neutrinos are expected to dominate the $\nu_e$ spectrum at energies above $\sim30\,\mathrm{TeV}$~\cite{beacom}.
The production of $\nu_\tau$ is expected to be negligible.

Fermi acceleration of charged particles in magnetic shocks followed by collisions with matter or radiation between the source and the Earth naturally leads to an energy 
spectrum for extraterrestrial neutrinos  that is harder than that for atmospheric neutrinos,  typically close to $dN_\nu/dE_\nu \propto E_\nu^{-2}$.
This allows diffuse extraterrestrial neutrinos to be visible as a hard component to the observed spectrum.
The ratio of the $\nu_e:\nu_\mu:\nu_\tau$ flux in a single astrophysical source depends on the neutrino energy~\cite{kashti05}. 
At moderate (high) energies, the neutrino flavor flux ratio behaves like the one from a pure pion (muon-damped) source, leading to an observed 
$1:1:1$ (1:1.8:1.8) ratio at the Earth after taking into account neutrino oscillations. The energy at which a flavor ratio transition occurs
thus depends on the properties of the source~\cite{kashti05}. 
The neutrino flux  is not known, although it is expected to be below the Waxman-Bahcall bound~\cite{WB}.

Previous searches for a diffuse flux have been performed with muon neutrinos \cite{ICCRspectrum}, and with cascades~\cite{ICCascades,OxanaPaper}.   
Cascades are the particle showers (electromagnetic and hadronic) initiated by charged current interactions of $\nu_e$ and $\nu_\tau$ and the neutral-current neutrino interactions of all three flavors in a medium.
In the charged-current interactions, an average of $80$\%  of the (high energy) neutrino energy goes into the produced lepton~\cite{gandhi}.
For $\nu_e$, this leads to an electromagnetic shower, while for $\nu_\tau$ the character of the lepton-induced shower depends on the $\tau$ decay mode.  
The remainder of the energy is transferred to the target nucleon, producing a hadronic cascade.  
In the neutral-current interactions, the neutrino transfers a fraction of its energy to the target nucleon producing only a hadronic cascade.
A typical cascade deposits its electromagnetic energy in a thin cylinder about 30 cm in radius and 5 m in length. Hadronic energy is deposited over a larger volume, about 11 m long and 75 cm in diameter.
IceCube observes the Cherenkov radiation produced by the charged secondary particles from neutrino-nucleon interactions through an optical sensor array. While a neutrino-induced muon has a track-like signature in IceCube, a cascade event looks effectively like a point source of Cherenkov light in the detector.

For diffuse searches, cascades from all flavor $\nu$ interactions have two advantages over tracks from $\nu_\mu$ interactions,
despite their inherently poor angular resolution compared to muon tracks.
The first is that the background from atmospheric neutrinos is  lower than for  $\nu_\mu$.  Second, because of their short shower length, the cascades are well-contained in the detector, 
with a Cherenkov light output proportional to the shower energy, so the shower energy is well measured. The detector acts as a calorimeter. 
Since the energy spectrum of extraterrestrial neutrinos is expected to be harder than the atmospheric neutrino spectrum, searching for a break in the energy spectrum with 
cascades is easier than with muons, both due to the expected break being at a lower energy in the cascade channel than the muon channel (a consequence of  
lower fluxes of atmospheric $\nu_{e}$ than $\nu_{\mu}$), and better intrinsic energy resolution of cascades over muons.

This paper reports on searches for diffuse extraterrestrial and for atmospheric neutrino-induced cascades using 
$257$ days (livetime) of data  collected in the year 2007-2008 with a partially completed IceCube detector consisting of 22 of the planned 86 strings. 
The IceCube detector and data sample are described in section~\ref{sec:detector}.   Section~\ref{sec:analysis} describes the analysis.  Results are given in section~\ref{sec:results}, and  
a summary follows in section~\ref{sec:summary}.

\begin{figure}
\centering
\hspace*{-0.35cm}\includegraphics[width=0.3\textwidth]{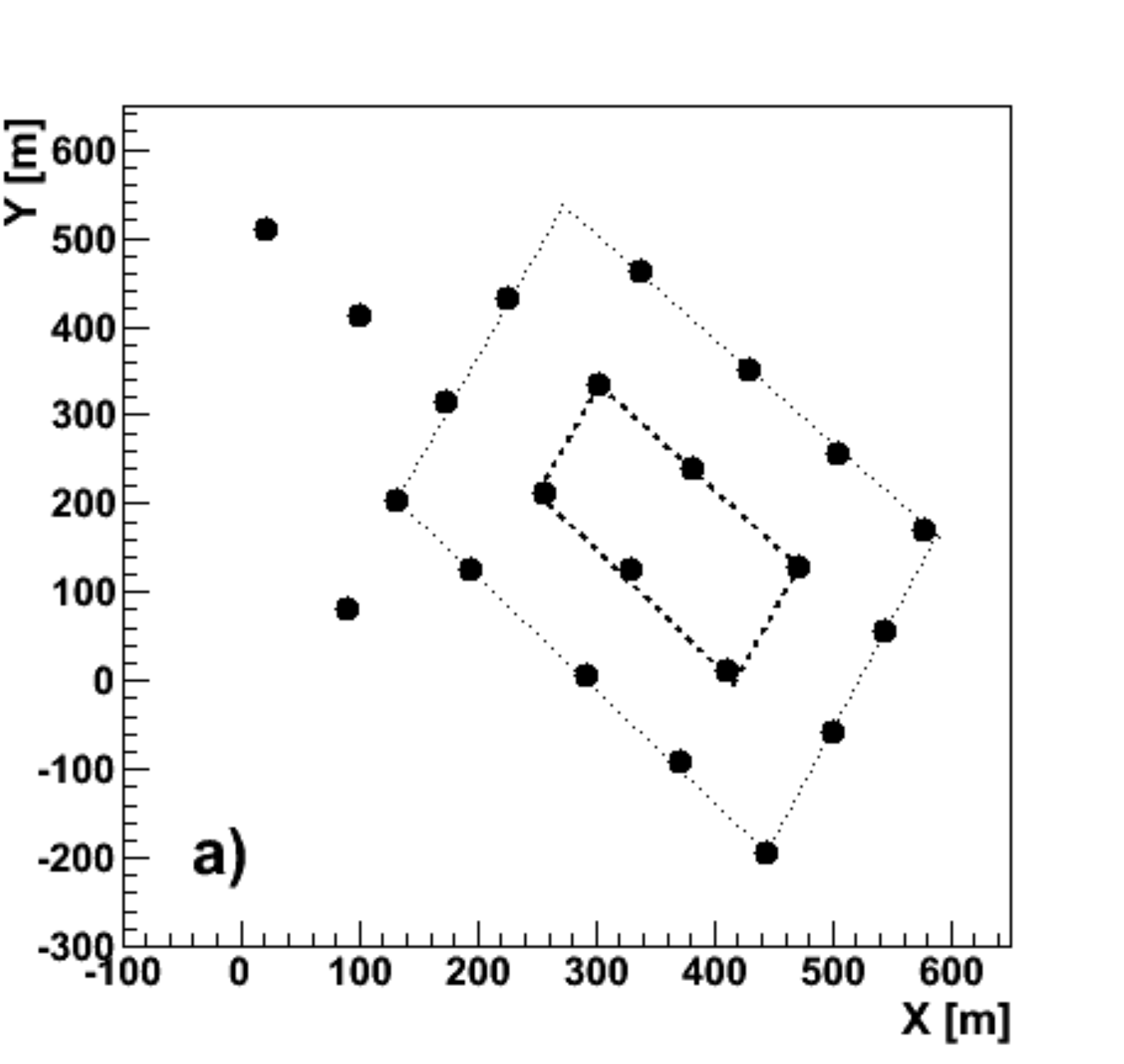}
\includegraphics[width=0.34\textwidth]{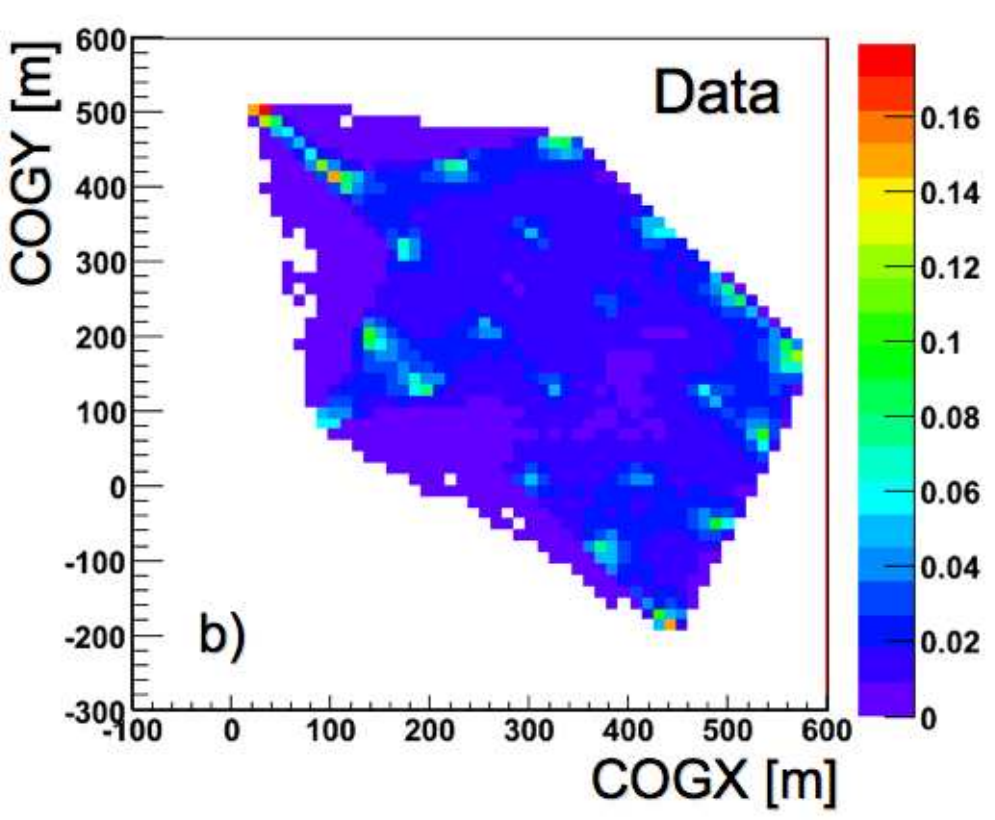}
\caption{(Color online) a)  
The filled circles show the positions of the strings in the $x-y$ (horizontal) plane for the $22$-string detector configuration.
The lines show the boundaries of the fiducial volume and are described in the text.
b)  The reconstructed center-of-gravity position $x$ (COGX) versus $y$ (COGY)  for events passing the cascade online filter. 
The right axis shows the rate [Hz].}
\label{fig:fiducial}
\end{figure}

\section{ The IceCube Detector and Data Sample}\label{sec:detector}
The IceCube detector is composed of vertical strings of optical sensors which are deployed in the glacial ice at the South Pole.    The sensors detect  Cherenkov radiation from charged particles produced in neutrino interactions.   
The strings are deployed on a 125 m triangular grid.  Each string contains 60 digital optical modules (DOMs), mounted between 1450 and 2450 m below the surface (17 m spacing).  
Each DOM contains a 10-inch photomultiplier tube (PMT)~\cite{PMTpaper}, and a data acquisition system in a pressure vessel.    The DOMs are most sensitive to photons
with wavelengths 300 to 600 nm.  The Hamamatsu R7081-02 PMTs have a peak quantum efficiency of $\sim25$\% 
and are operated  at a typical gain of $10^7$.
The PMT output is linear within 10\% up to a current of 50 mA; at a gain of $10^7$, this corresponds to 31 detected photoelectrons per nanosecond. 
The DOM and string performance is described elsewhere~\cite{IC1performance,SORMA}.

The data acquisition system records the arrival times of the detected photoelectrons. It uses two waveform digitization systems which record the arrival time 
with a time accuracy of about 2 ns and a wide dynamic range~\cite{MBpaper,IC1performance}.  One system records data at 300 megasamples/second (MSPS) for 400 ns 
after the first photon is detected with 14 bits of dynamic range.  The second only has 10 bits of dynamic range but records data for 6.4 $\mu$s at 40 MSPS.  
The dead-time of the system is less than $1\%$.  Each DOM is activated (launched) when a single photoelectron is detected, and the data are sent to the surface 
when two adjacent (nearest or next-to-nearest) neighbors record a hit within 1 $\mu$s.

The data were collected between May 2007 and April 2008 when IceCube consisted of 22 active strings with 1320 DOMs.
The detector configuration is shown schematically in Figure~\ref{fig:fiducial}a.
The point (0,0,0) is the center of the complete $86$-string IceCube detector.
The main physics trigger was a ``simple  multiplicity trigger" 
that required photon signals 
from at least 8 DOMs, with the additional requirement of accompanying hits in any of the two neighboring DOMs on a string, 
each above a threshold of 1/6 single photoelectron and within a $5$ $\mu$s coincidence window.
The average trigger rate was $\sim550$ Hz, 
driven by atmospheric muons, and exhibited about $\pm10\%$ seasonal variation.
Online filters were employed to preselect a data sample for satellite transmission, that was used for analysis.

The online filter relevant to this paper used two first-guess reconstruction algorithms~\cite{cascade-reco}.
One algorithm assumed that all hits can be projected onto a line consistent with a particle traveling at a specific velocity.
The second algorithm quantified the sphericity of the hit topology.
It used the center-of-gravity (COG), defined as the photon signal amplitude weighted mean of all hit DOM positions, 
as a first-guess vertex position. 
The response of the algorithms was studied with Monte Carlo simulation.  
Online selection criteria were developed using these simulations to reject 
97.5\% of the background events, while retaining 70\% of the cascade signal events.  
The average rate after this filtering was 19~Hz.

\section{Analysis}\label{sec:analysis}
Even after online filtering, the data are dominated by atmospheric muons produced in interactions of cosmic rays with nuclei in the Earth's atmosphere. 
It is thus necessary to develop stringent offline selections to reject this background while retaining cascade signal events.
Two such analyses were performed independently.  One analysis focused on cascade events induced by high energy extraterrestrial neutrino interactions, while the other searched for lower energy atmospheric 
neutrino-induced cascade signals.
The development of the selections for both analyses relies on using 10\% of the recorded data 
while keeping the remaining 90\% ``blinded". 
When the selections have been fixed, the 10\% sample is discarded and the remaining data is unblinded.
The physics results were evaluated from the remaining 90\% of the recorded data only after the analysis selections were finalized and were thus free of statistical bias.
In addition, extensive Monte Carlo simulations of background and signal events, as well as the IceCube detector response, were used.
These simulations used importance sampling and weighting techniques to overcome computing limitations and are described in section~\ref{sec:simulation}.
This is followed by a description of the two analyses in section~\ref{sec:selections}. Systematic uncertainties are described in section~\ref{sec:systematics}.

\subsection{Monte Carlo Simulations}\label{sec:simulation}
The Monte Carlo generator ANIS~\cite{anis} was used to generate neutrinos of all flavors in the energy range from $\sim10$ GeV to $1$ EeV 
at the surface of the Earth and to simulate subsequent neutral and charged current interactions in the Earth.
The neutrino spectra were generated with energy distributions of $E^{-2}$ and $E^{-1}$, and were re-weighted to conventional and prompt atmospheric neutrino flux predictions or 
to an astrophysical neutrino flux of $E^{2}\Phi = 1 \times 10^{-6}$  ${\rm{GeV \cdot cm^{-2} \cdot s^{-1} \cdot sr^{-1}}}$, depending on the analysis goal.
The CTEQ5 parton distribution functions~\cite{cteq} were used in evaluating the neutrino cross sections. 
The Preliminary Reference Earth Model~\cite{refmodel} was used as the density profile to simulate high energy neutrino absorption.
The simulations included the Glashow resonance $\bar{\nu}_{e} + e^- \rightarrow  W^{-} \rightarrow X$ for $\bar{\nu}_{e}$ energies near $6.3$ PeV~\cite{glashow}.

CORSIKA~\cite{corsika} was used to simulate air showers from cosmic ray collisions in the Earth's atmosphere and obtain a sample of background atmospheric muons.
The cosmic ray spectrum was based on the H\"orandel polygonato model~\cite{Hoerandel}, and primaries up to iron were simulated in the energy range from $600$ GeV up to $10^{11}$ GeV.
The SIBYLL parameterization~\cite{sibyll} was used in CORSIKA to model hadronic interactions.
The effects of the Earth's magnetic field were included in the simulations.
By overlaying independent CORSIKA showers, we simulated the coincident muon background, which comes from multiple cosmic ray showers triggering the detector  within 
its readout window.

High energy background muons are likely to incur large radiative energy losses, which may mimic cascade signals.  To obtain an adequate sample of such high energy background muons in the simulations, it was necessary to impose additional selections in the CORSIKA event generations.  We have imposed threshold selection criteria on the primary cosmic ray energy and on the energy of individual muons, as was done in the AMANDA analysis~\cite{OxanaPaper}, to increase the effective livetime and Monte Carlo statistics for cosmic rays in the energy range from $\mathcal{O}(10^2)$ TeV to $\mathcal{O}(1)$ PeV.
The effective livetimes of these simulation samples are given, together with the threshold values, in Table~\ref{tbl:simulations} for representative cosmic ray energies.
The high-energy sample was used only to assess background.

\begin{table}[t]
	\caption{The effective livetimes of the CORSIKA Monte Carlo simulation samples, generated with standard energy thresholds of  600\,GeV for primary cosmic rays and 273\,GeV for secondary muons and with high-energy thresholds of 40\,TeV and 5\,TeV, respectively.}\label{tbl:simulations}
	\centering
	\begin{tabular}{c@{~~~~}c@{~~~~}c}
	\hline\hline
		Primary Cosmic Ray Energy & \multicolumn{2}{c}{Monte Carlo sample} \\ \cline{2-3}
		& standard & high-energy \\
		& [days] & [days] \\
	\hline
		\begin{tabular}{r}
			10 TeV \\ 100 TeV \\ 1 PeV \\ 10 PeV
		\end{tabular} &
		\begin{tabular}{r}
			1 \\ 10 \\ 100 \\ 1000
		\end{tabular} &
		\begin{tabular}{r}
			-- \\ 50 \\ 500 \\ 5000
		\end{tabular} \\
	\hline\hline
	\end{tabular}
\end{table}
A Muon Monte Carlo simulator~\cite{MMC} was used to propagate secondary muons through the ice.
It simulates muon stochastic energy losses from ionization, bremsstrahlung, photo-nuclear interactions, and pair production.

A Cascade Monte Carlo~\cite{CMC} was used to simulate  the longitudinal development of electromagnetic and hadronic cascades.  
It also accounts for the lower Cherenkov light output from hadronic cascades compared to that from electromagnetic cascades.

Cherenkov light emission and subsequent photon propagation through ice was simulated with the Photonics simulation package~\cite{photonics}. In these simulations, the optical properties of the ice were described by a ``calibrated ice model".
This model was  constructed from extensive AMANDA measurements of light propagation in South Pole ice made with artificial in-situ light sources (pulsed and steady LED sources and nitrogen lasers)~\cite{IcePropertiesPaper}. 
These measurements were largely decoupled from light source and detector characteristics by using timing information of detected single photoelectrons. 
They determined the relevant wavelength and depth dependences of the optical scattering and absorption lengths, down to the deepest AMANDA depths. Using the fact that scattering and absorption are highly correlated with the concentration of insoluble dust particles in the ice, the model was extrapolated to the greater IceCube depths (from $2100\,{\mathrm{m}}$ down to bedrock at about $2800\,{\mathrm{m}}$) with dust concentration data from an Antarctic ice core~\cite{icecore}, 
which were scaled to fit the measured scattering and absorption parameters at AMANDA depths. The ice properties are thus less understood at depths greater than 
$2100\,{\mathrm{m}}$.

PMT response simulators were used for each DOM in the detector to relate light input and current output.  The simulated currents were then propagated through response simulators of the digitization electronics, local coincidence signaling, and event triggering.
The same processing and filtering was applied to simulated and recorded data.

\subsection{Event selections}\label{sec:selections}
After online filtering and transfer, the data were passed through several maximum-likelihood based reconstruction algorithms~\cite{track-reco,cascade-reco}
in order to suppress muon background while retaining signal events.
The algorithms assumed a single track, two track, and a point-like cascade hypothesis and obtained the corresponding vertex positions and times for each event.
Following the reconstruction, a sequence of selection criteria was applied to the data.  The retained data are identified by different levels, starting from Level-3 (Level-1 refers to triggered online data and Level-2 to the data before offline reconstruction).

At Level-3 only those events were retained for which the single muon track reconstruction found a zenith angle greater than $73^{\circ}$.  
In addition, a requirement was imposed on the difference of the log-likelihood, $\mathcal{L} $, 
of the cascade and track reconstructions for each event,
$\mathcal{L}(\mathrm{cascade}) - \mathcal{L}(\mathrm{track})> -16.2$,
to preferentially select cascade-like events.
The selection criteria were chosen so that the signal efficiency was about 80\%
in both the atmospheric and extraterrestrial neutrino searches and the muon background rate was reduced by a factor of $\sim5$,
with respect to the online filter.

 Both analyses used the same cascade energy reconstruction algorithm~\cite{MDA} which was applied to events passing Level-3. 
 This analytical cascade-energy  calculation takes into account the variation of ice optical properties with depth: 
 \begin{equation} 
 E_{\rm{reco}}/{\rm{GeV}} =  \frac{\sum{n_{\rm{pe}}}}{\sum{\mu_{0}\left( \vec{r}_{\rm{v}},\vec{r}_{\rm{DOM}} \right)}}.
 \label{eq:energy}
 \end{equation}
 The ${n_{\rm{pe}}}$ is the observed number of photoelectrons in a DOM and $\mu_{0}\left( \vec{r}_{\rm{v}},\vec{r}_{\rm{DOM}} \right)$ is the number of photoelectrons expected at a given DOM position $\vec{r}_{\rm{DOM}}$ from 
 a  $1\,\mathrm{GeV}$ cascade with  a  vertex position  $\vec{r}_{\rm{v}}$.  The $\mu_{0}$ is taken from a table generated by the Photonics Monte Carlo package~\cite{photonics}. 
 In Eq.~\ref{eq:energy} the noise term has been neglected.  It biases the reconstructed energy towards lower values  and thus worsens the performance of the cascade energy reconstruction.

The subsequent selections, described below, were optimized separately for the two analyses.

\subsubsection*{Extraterrestrial event selections}
Energetic bremsstrahlung from muon tracks outside the detector and muon tracks that intersect only part of the detector can mimic an uncontained cascade signal.
To reject this type of background, events were required to have a topology consistent with a cascade signal that originated inside the IceCube instrumented volume at Level-4.
Specifically, the four earliest hits in the event were required to be inside the fiducial volume of the detector in the horizontal $x$-$y$ coordinates, as depicted in Figure~\ref{fig:fiducial}a (dotted lines).  
Events were also rejected if any of the four earliest hits occurred in the eight topmost DOMs.
Approximately 5\% of the data and 7\% of the muon background events from the CORSIKA Monte Carlo remained with respect 
to the previous selection level. The difference between absolute rates in the data and the CORSIKA Monte Carlo is addressed in section~\ref{sec:systematics}.
ANIS Monte Carlo simulations show that approximately $13\%$ of the signal was retained at Level-4.

At Level-5, events were required to have hits in at least 20 DOMs.  To reject events with multiple muon tracks within the IceCube read-out window, an upper limit of $5$\,$\mu$s was set on the event duration, 
defined as the time difference between the last and the earliest hit in the event.  A two-track reconstruction was performed on the retained events and the 
tracks were required to coincide to within $1\,\mu\mathrm{s}$.
The center-of-gravity of the hits and the vertex coordinates from the cascade reconstruction algorithm were required to coincide in $x$ and $y$ to within $60\,\mathrm{m}$.
Monte Carlo simulation studies indicate that these selections reduce the coincident muon rate to less than 20\% of the $8\,\mathrm{mHz}$ expected background muon rate at level 5.

\begin{figure}[ht]
\centering
\includegraphics[width=0.41\textwidth]{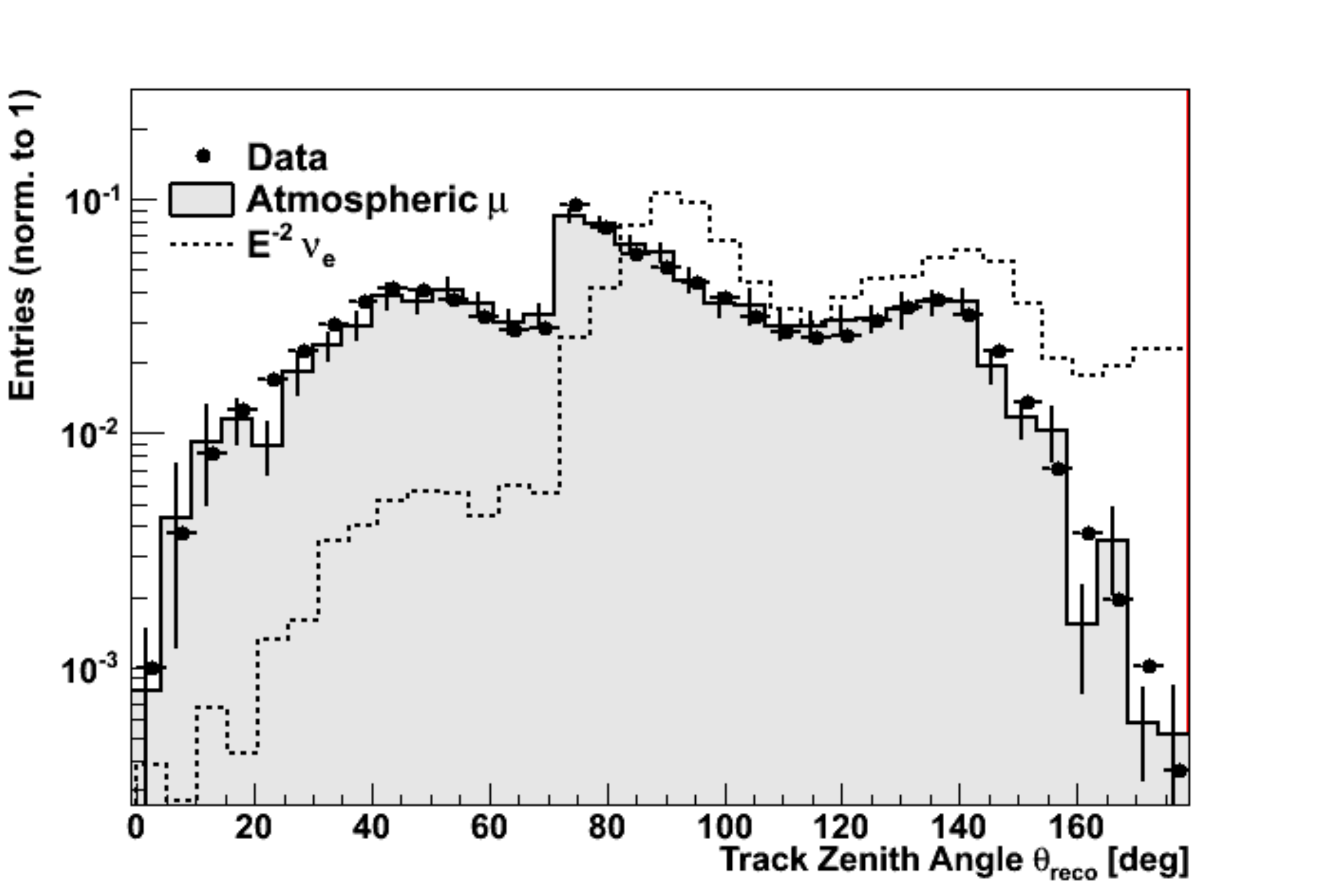}
\vspace*{-0.2cm}
\includegraphics[width=0.41\textwidth]{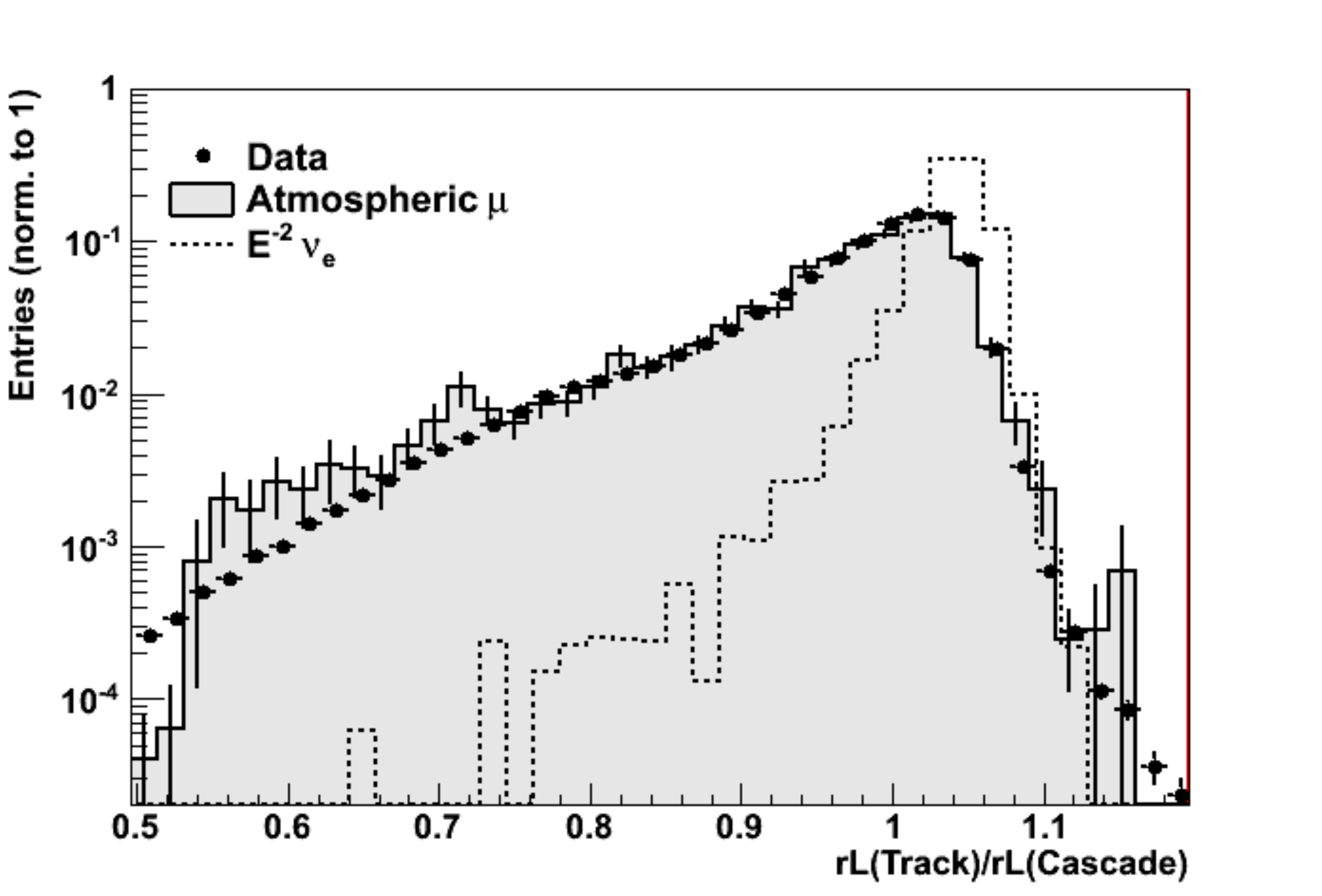}
\caption{Normalized distributions of  (top)  zenith angle and (bottom) reduced log likelihood ratios
after Level-5 event selections for the data (filled circles), muon background Monte Carlo (continuous histogram) and 
signal neutrino (dashed histogram). }
\label{fig:zenith}
\end{figure}

The smaller event sample allowed for the use of more CPU-intensive reconstruction algorithms.
At Level-6, the single-track reconstruction was iterated 32 times with randomly selected seed tracks 
to ensure that the final reconstructed tracks did not originate from local minima.
The zenith angle distribution is shown in the top part of Fig.~\ref{fig:zenith}. 
The selection on zenith angle was refined to reject events with a reevaluated single-track zenith angle smaller than $69^{\circ}$.
The comparison of the event reconstruction probabilities with a single-track and a cascade hypothesis was revisited as well.
Only those events for which the ratio of the reduced log-likelihood, 
$\mathrm{r}\mathcal{L}$, for the event reconstruction under a single-track and under a cascade hypothesis was greater than 0.95 were retained for further analysis. 
The $\mathrm{r}\mathcal{L}$ is defined as the negative log-likelihood normalized by the number of degrees of freedom.
The normalized distribution of the track and cascade reduced likelihood ratio is shown in the bottom part of Fig.~\ref{fig:zenith}. 
The Monte Carlo describes the shapes of experimental data distributions very well. 
The rates  observed in the data and obtained from simulations are given in Table~\ref{tbl:rates} and discussed 
in section~\ref{sec:systematics}.

\begin{figure}
\centering
\includegraphics[width=0.42\textwidth]{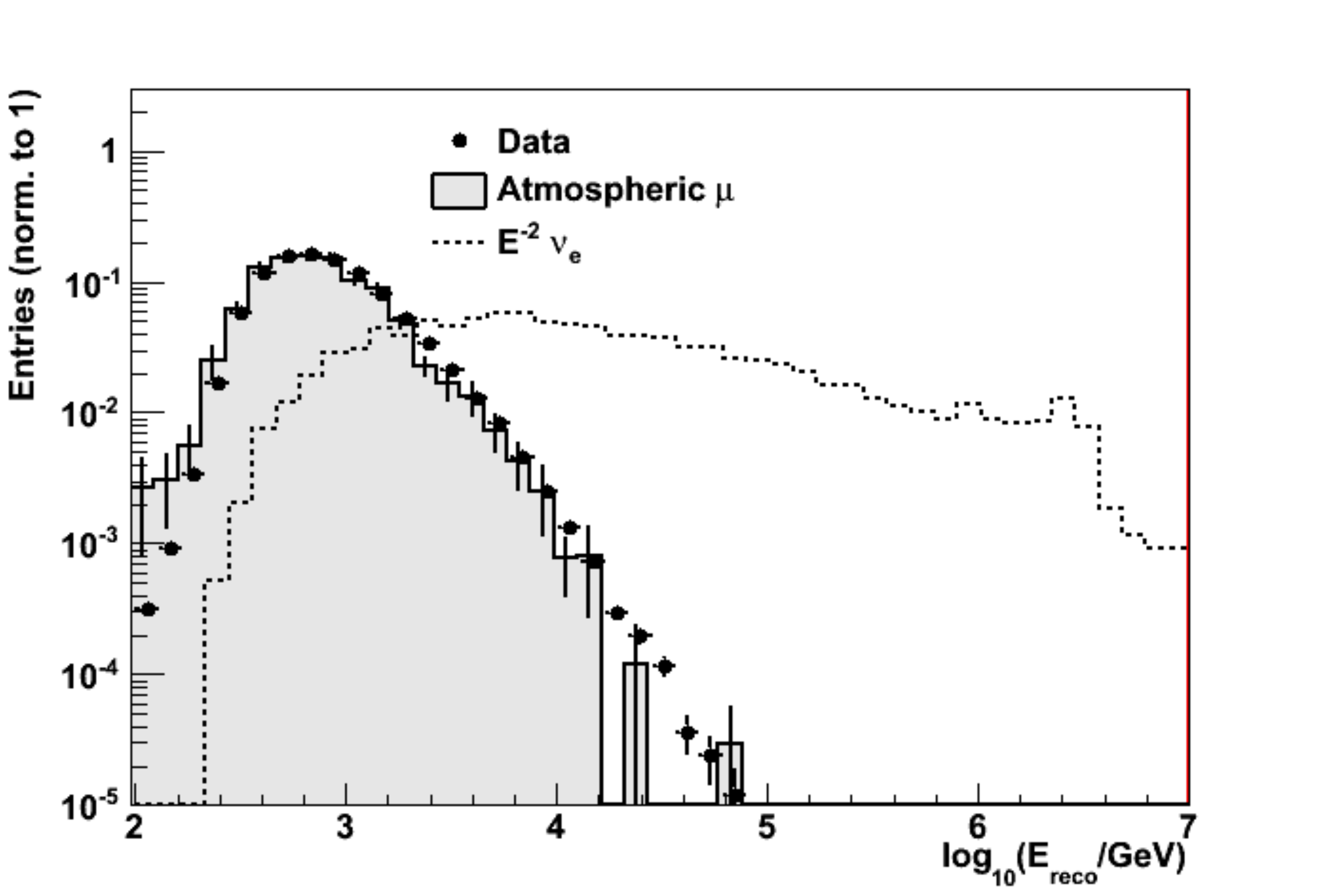}
\vspace*{-0.2cm}
\includegraphics[width=0.42\textwidth]{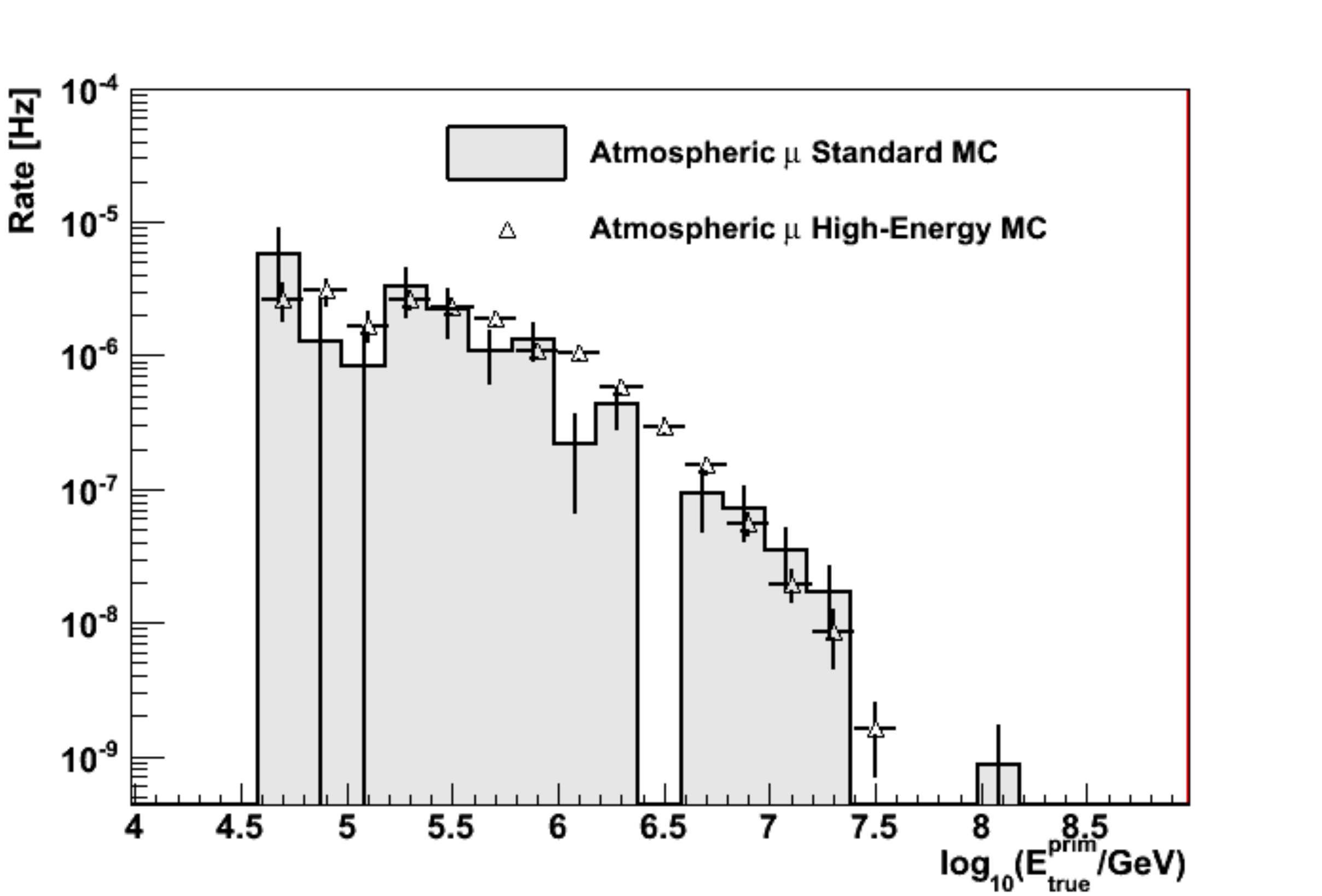}
\caption{(top) Normalized distribution of the reconstructed cascade energy after level-6 event selections
for the data  (filled circles), muon background Monte Carlo (continuous histogram) and 
signal neutrino (dashed histogram).  (bottom) Absolute event rate versus primary Cosmic Ray energy after  level-6 event selections  with 
an additional selection criterium on the reconstructed cascade energy of 6.3 TeV for the standard Corsika Monte Carlo (continuous histogram)
and high-energy optimized Corsika Monte Carlo (open triangles).}
\label{fig:energy}
\end{figure}

\begin{figure*}
\begin{minipage}[h]{0.99\linewidth} 
\includegraphics[width=0.32\textwidth]{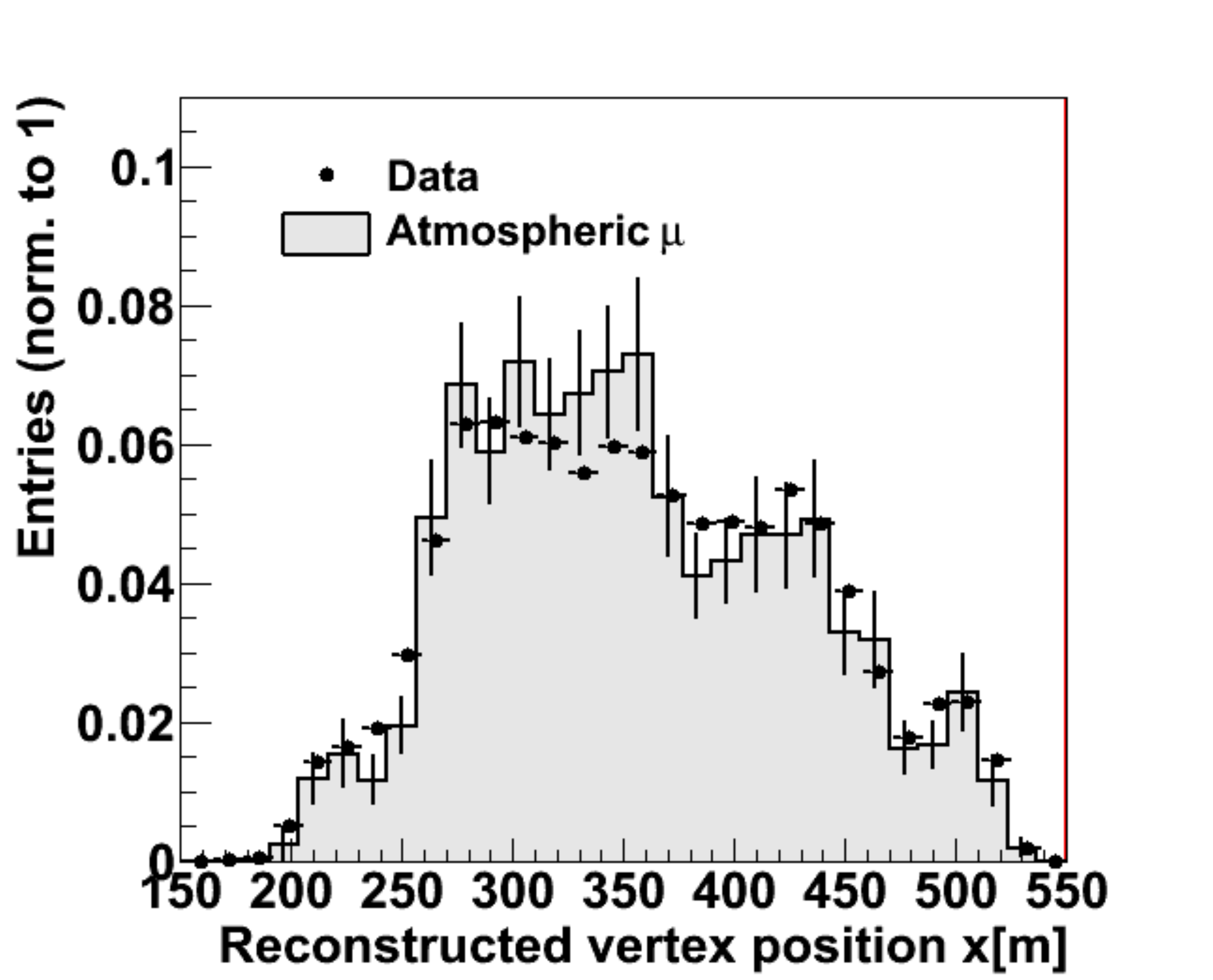}
\includegraphics[width=0.32\textwidth]{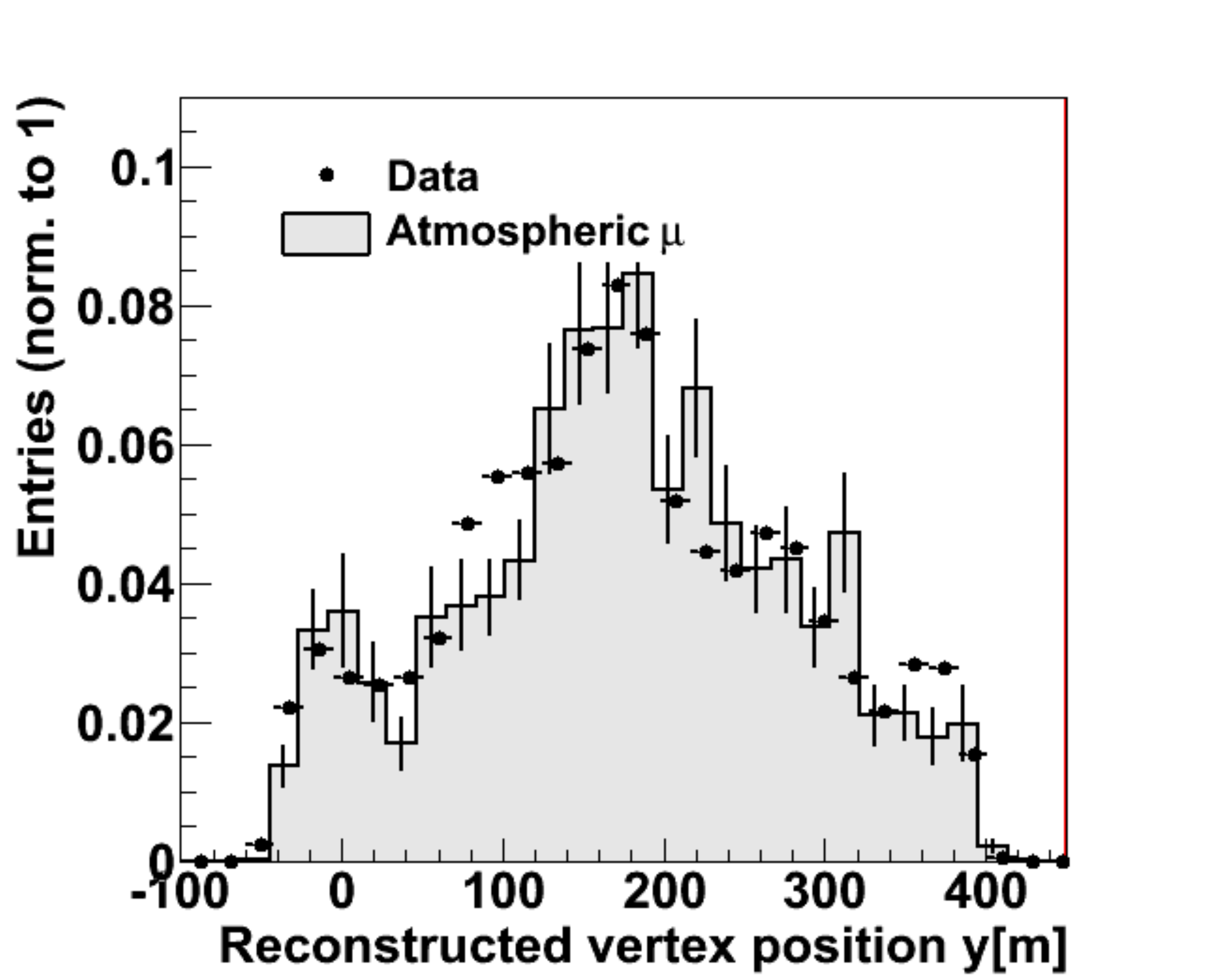}
\includegraphics[width=0.32\textwidth]{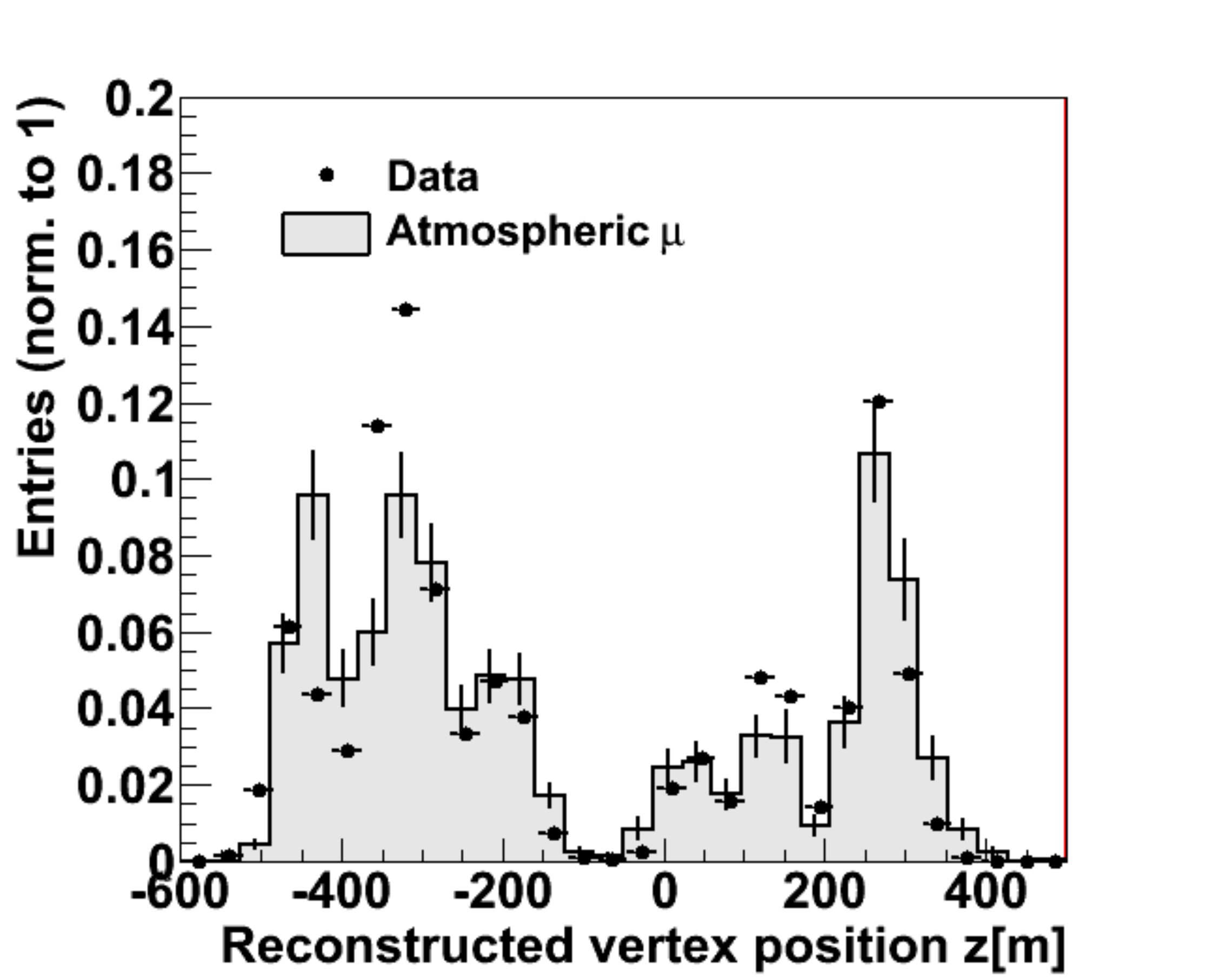}
\caption{Normalized distributions of  cascade reconstructed vertex position: $x$ component (left panel), $y$ component (middle panel)  and $z$ component (right panel) 
after level-6 event selections for the data (filled circles), standard muon background Monte Carlo (continuous histogram).}
\label{fig:vertex}
\end{minipage}
\end{figure*}

Figures~\ref{fig:energy} (top) and ~\ref{fig:vertex} show comparisons between the Monte Carlo and the data for the reconstructed energy, and the cascade reconstructed vertex positions 
after Level-6 selections.  
The shapes agree very well, except  vertex position $z$ at the largest depths of the detector where the ice properties are not well described by our simulations. 
Figure~\ref{fig:energy} (bottom) shows the absolute event rate as a function of primary Cosmic Ray energy for the standard and high-energy optimized Corsika Monte Carlo
after level-6 event selections with an additional selection criterium on the reconstructed cascade energy of 6.3 TeV.
Since the two spectra are consistent within uncertainties, at Level-7 we use the high-energy optimized Corsika Monte Carlo.

\begin{figure}
\begin{minipage}[b]{0.999\linewidth}
\centering
\includegraphics[width=0.99\textwidth]{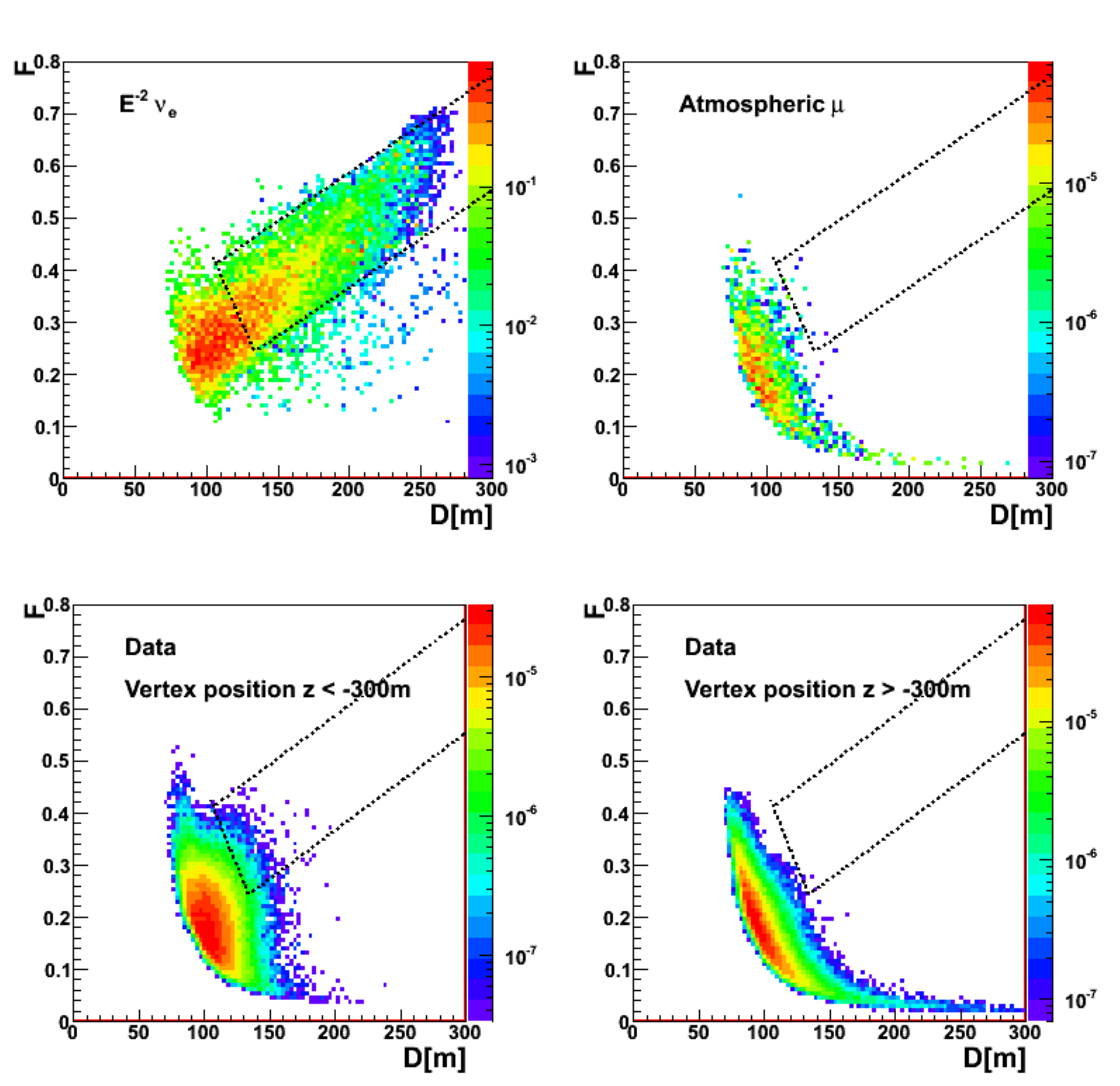}
\caption{(Color online) The fill-ratio, $F$, versus the distance, $D$, as defined in the text for (top-left panel) the Monte Carlo simulated signal sample, (top-right panel) the Monte Carlo 
simulated muon background sample, and the data at the bottom part of the detector (bottom-left panel) and at the top part of the detector (bottom-right panel).   
The right axis shows the rate [Hz]. The dashed lines show the selection boundaries applied at Level-7 in the analysis. }
\label{fig:FD}
\end{minipage}
\end{figure}

In order to further distinguish between signal-like and muon-like events, at Level-7 we used the multiplicity and spatial distribution of the hit DOMs.
For every event we calculated  the distances between each hit DOM  and the reconstructed cascade vertex. 
Event-by-event, the mean distance, $D$, was evaluated and used as the half-radius in determining fill-ratio $F$, defined as a fraction of hit DOMs within a sphere centered 
on the cascade vertex for the event.  
Monte Carlo studies of signal and background exhibit different correlations of $D$ and $F$, as shown in Figure~\ref{fig:FD}. 
For signal-like events, the  average $F$ value (sphere density) increases as a function of $D$, while the opposite is true for the   
single muon events. Most of coincident muon events have very small values of $F$, independent of $D$.
Figure~\ref{fig:FD} also shows the distribution of $D$ and $F$ in the data and Monte Carlo simulations.
The background Monte Carlo is in good agreement with the data for the events with the cascade reconstructed vertex position $z>-300\,{\mathrm{m}}$
as shown in Figure~\ref{fig:FD} (bottom-right panel).
This is not the case for  the events with the cascade reconstructed vertex position $z<-300\,{\mathrm{m}}$, 
where the  ice  is less well understood and makes some muons look like non-contained cascades (spherical shape and high DOM multiplicity).
The muon background events were removed  by increasing 
a threshold on the minimum number of hit DOMs to 60 and imposing a  threshold  
$E_\mathrm{reco} > 16\,\mathrm{TeV}$ on the reconstructed cascade energy.  
These selections also suppress contributions from atmospheric neutrinos from pion and kaon decays.
After all selections, four events in the muon background simulation samples and no events in the 10\% of the data remained~\cite{JK}.

\begin{table*}
\begin{minipage}[b]{0.9\linewidth}
\centering
\caption{Event rates at different selection levels for the extraterrestrial and atmospheric analyses, described in the text, for the data and for Monte Carlo simulations of atmospheric background muons and of atmospheric and astrophysical neutrinos.  An astrophysical neutrino flux of $E^{2} \Phi_\mathrm{model} = 1.0 \times 10^{-6} {\rm{ GeV \cdot cm^{-2} \cdot s^{-1} \cdot sr^{-1}  }}$ was used  for all $\nu$ flavors. The statistical uncertainties in the event rates are smaller than 10\%. }\label{tbl:rates}
\begin{tabular}{cc@{}c@{}c@{}c@{}c@{}c@{}c@{}c@{}c}
\hline\hline \\[-2.0ex]
Selection level & Observed Rate & \multicolumn{8}{c}{Simulated Rate} \\
  & 90\% Data & $\mu$ & & $\nu_e^{\rm{atm}}$ & $\nu_\mu^{\rm{atm}}$ & & $E^{-2}~\nu_e$ & $E^{-2}~\nu_\mu$ & $E^{-2}~\nu_\tau$ \\

& [Hz]  & [Hz] & & [Hz] & [Hz]  & &  [Hz] & [Hz] & [Hz] \\
\hline \\[-2.0ex]  
3 & 3.7 & 3.0 & & 5.2$\times 10^{-5}$ & 2.9$\times 10^{-4}$ & & 6.5$\times 10^{-5}$ & 4.3$\times 10^{-5}$ & 9.0$\times 10^{-5}$ \\
\hline \\[-2.0ex]
 Extraterrestrial & & \multicolumn{4}{c}{Background} & & \multicolumn{3}{c}{Signal} \\ 
 \cline{3-6} \cline{8-10} \\[-2.0ex]
\begin{tabular}{c}

4 \\ 5 \\ 6 
\end{tabular} &
\begin{tabular}{l} 
2.6$\times 10^{-1}$ \\ 2.1$\times 10^{-2}$ \\ 1.1$\times 10^{-2}$ 
\end{tabular} &
\begin{tabular}{l} 
1.5$\times 10^{-1}$ \\ 8.3$\times 10^{-3}$ \\ 4.1$\times 10^{-3}$ 
\end{tabular} &
\begin{tabular}{c}
 \\ \\ \\

\end{tabular} &
\begin{tabular}{l} 
1.2$\times 10^{-5}$   \\  3.7$\times 10^{-6}$   \\  3.3$\times 10^{-6}$ 
\end{tabular} &
\begin{tabular}{l} 
7.5$\times 10^{-5}$ \\  3.4$\times 10^{-6}$ \\  2.9$\times 10^{-6}$ 
\end{tabular} &
\begin{tabular}{c}
 \\ \\ \\

\end{tabular} &
\begin{tabular}{l} 
1.2$\times 10^{-5}$ \\  8.6$\times 10^{-6}$  \\  8.2$\times 10^{-6}$ 
\end{tabular} &
\begin{tabular}{l} 
6.8$\times 10^{-6}$\\   3.1$\times 10^{-6}$  \\   2.8$\times 10^{-6}$  
\end{tabular} &
\begin{tabular}{l} 
1.4$\times 10^{-5}$  \\ 1.1$\times 10^{-5}$   \\  1.0$\times 10^{-5}$  
\end{tabular} \\
\hline \\[-2.0ex]
 Atmospheric & & Background & & \multicolumn{2}{c}{Signal} & & & \\
 \cline{3-3} \cline{5-6} \\[-2.0ex]
4 & 3.1$\times 10^{-3}$ & 1.8$\times 10^{-3}$ & & 1.3$\times 10^{-6}$ & 1.5$\times 10^{-6}$ & & - & - & - \\

\hline\hline
\end{tabular}
\end{minipage}
\end{table*}

\subsubsection*{Atmospheric Neutrino Event Selection}

The atmospheric neutrino analysis used artificial neural networks implemented in the ROOT TMVA package~\cite{TMVA} to reject background.  
The neural networks had two hidden layers, the first with $N+1$ neurons and the second with $N$ neurons, where $N$ is the number of discriminating variables.

\begin{figure}[b]
\centering
\includegraphics[width=0.4\textwidth]{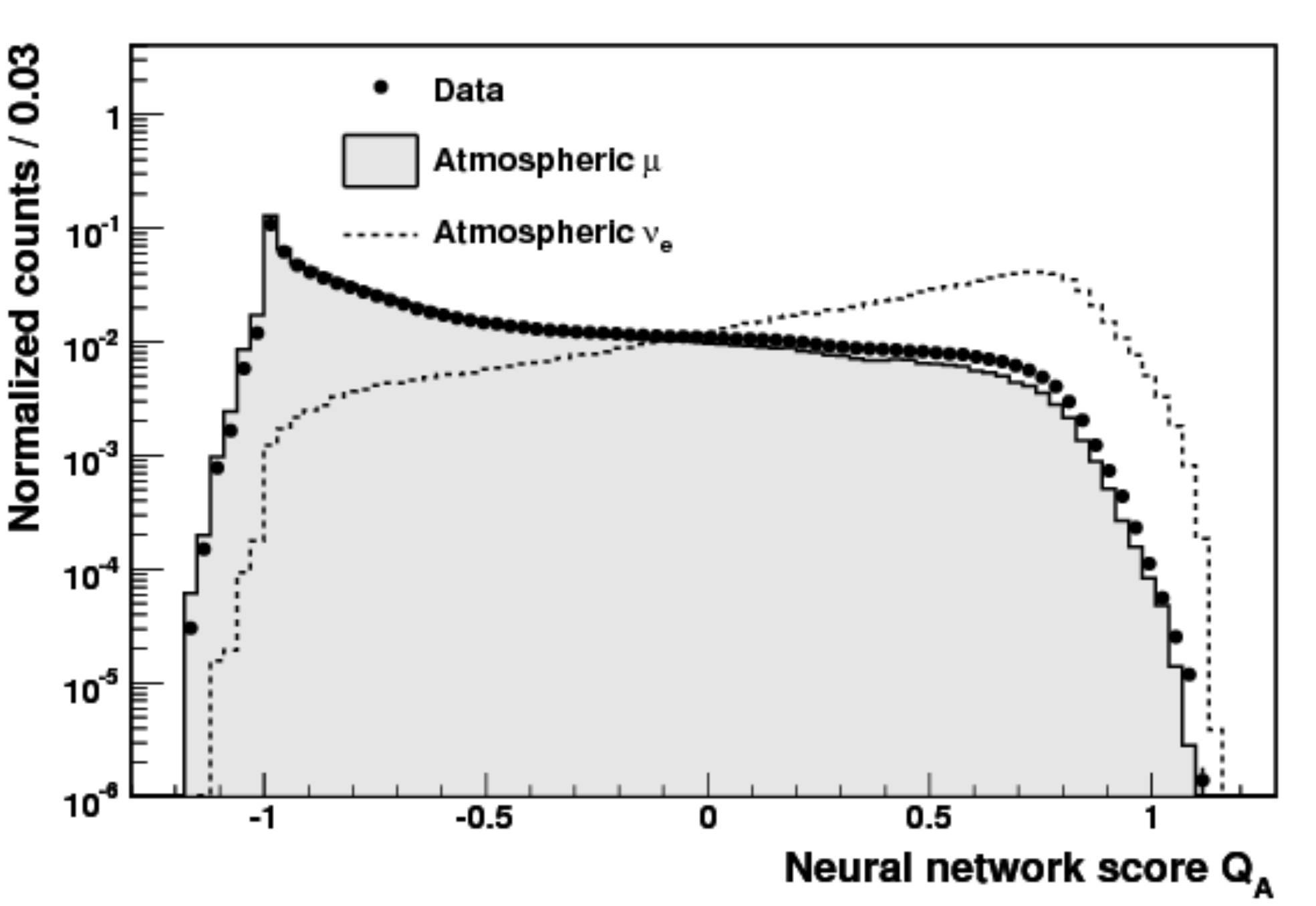}
\caption{Normalized distributions of the Level-4 neural network 
classification score.
}
\label{fig:mlplevel3}
\end{figure}

At Level-4, five input variables were used; the track zenith angle of the 32-fold iterative muon reconstruction, the reduced likelihood parameter from the cascade vertex reconstruction, the number of DOMs that register a single photoelectron divided by the total number of photoelectrons seen by all DOMs, the number of direct,  unscattered  photons assuming a point source at the reconstructed cascade vertex,
and the difference in $z$ vertex positions when the cascade vertex is reconstructed using only the earliest registered hits and using only the latest registered hits
for the so-called split cascade reconstructions.  
Figure~\ref{fig:mlplevel3} shows the classification score, 
${\rm{Q_A}}$, of the neural network trained from these variables.
Only events with
${\rm{Q_A}}> 0.4$ were retained.  In addition, the reconstructed cascade was required to have an energy larger than 2\,TeV and 
be contained within the detector by imposing an upper value of 1.4 on the parallelogram distance, $\alpha$, of its horizontal vertex position.
The six innermost strings span a parallelogram in $x$ and $y$, whose edges define $\alpha \equiv 1$.
The neighboring strings span a stretched parallelogram with $\alpha = 2$. Parallelograms of $\alpha = 1$ ($\alpha = 2$) are shown as dashed (dotted) lines in Fig.~\ref{fig:fiducial}a).
The containment selection criterium $\alpha<1.4$ thus requires the vertex to lie at least $75\,\mathrm{m}$ from the edge of the detector.
With these selections, background was reduced by a factor of $1700$ and $2.4\%$ of the signal events were retained.

\begin{figure*}
\centering
\includegraphics[width=0.32\textwidth]{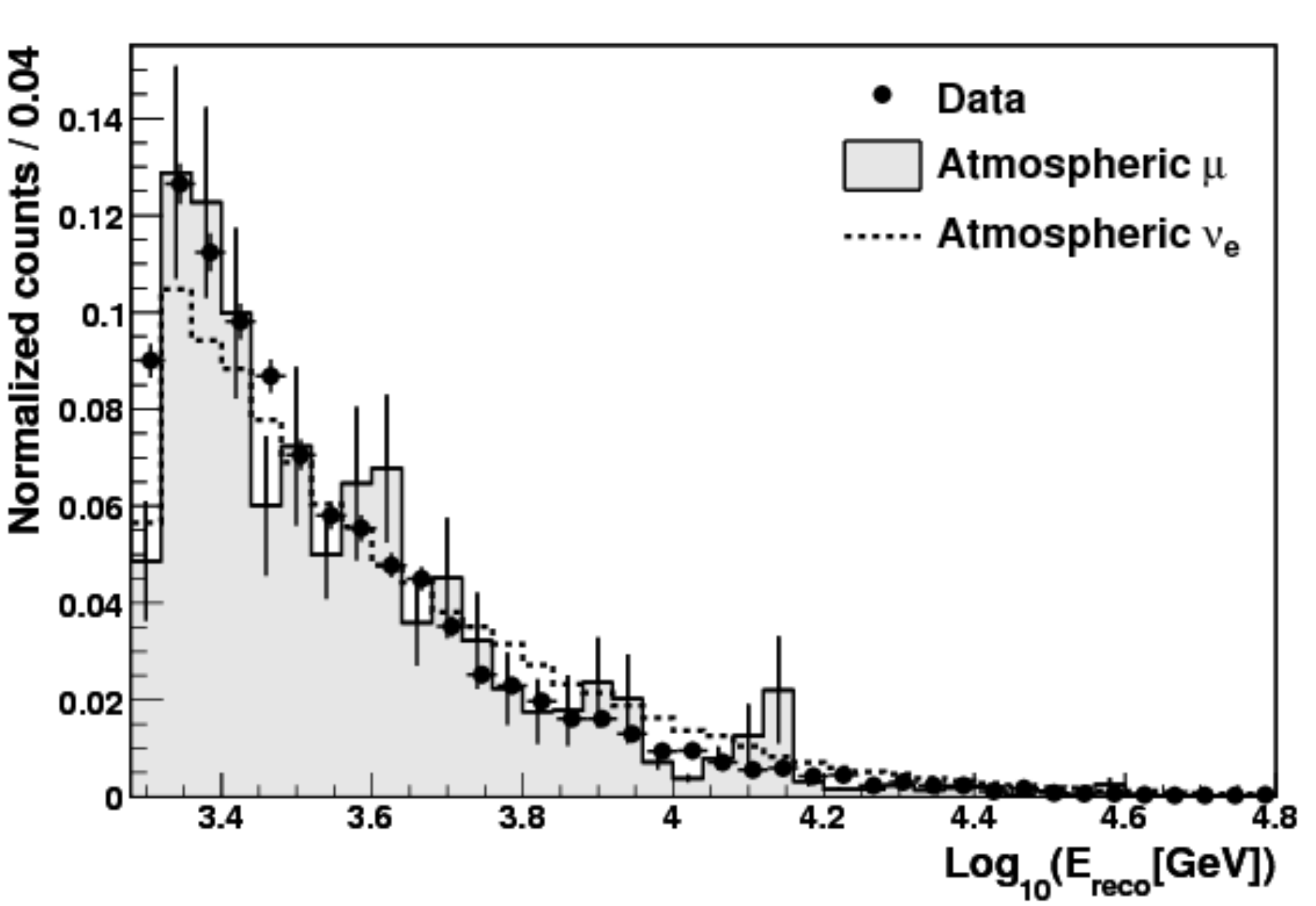}
\includegraphics[width=0.32\textwidth]{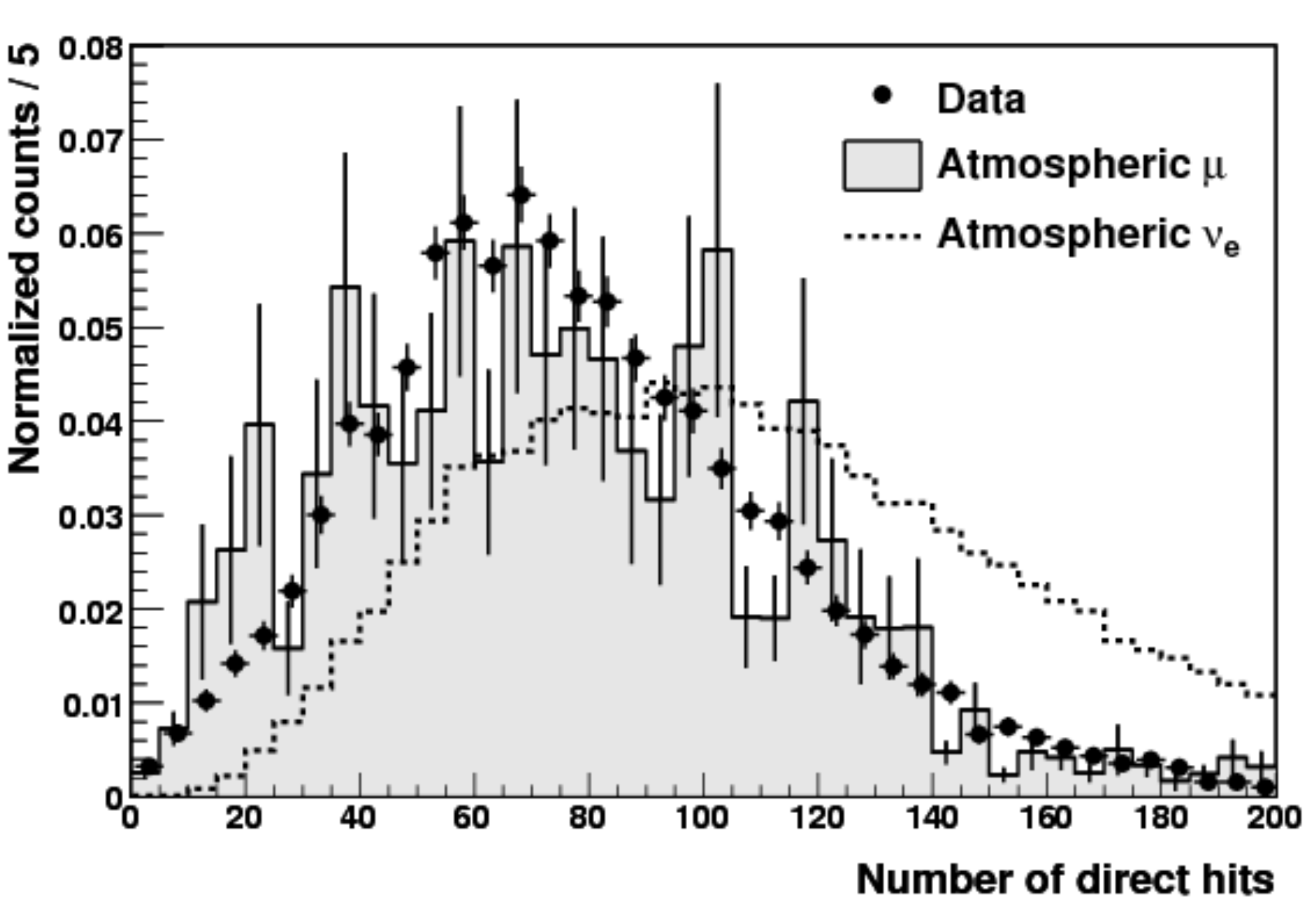}
\includegraphics[width=0.32\textwidth]{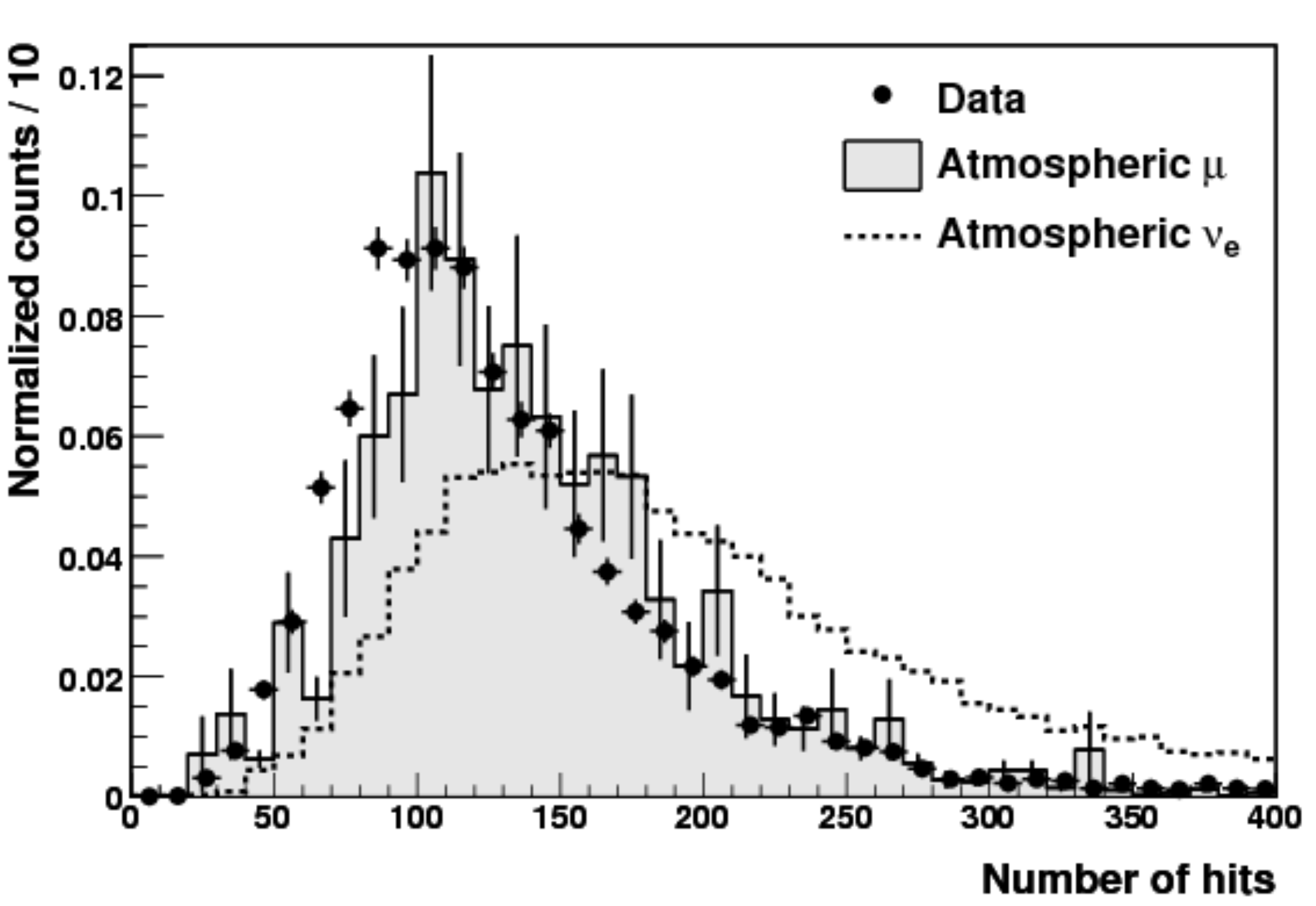}
\caption{Normalized distributions (after Level-4 atmospheric selections)  of (left panel) reconstructed energy (middle panel) number of direct hits  (right panel) total number of hits.
}
\label{fig:distlevel4a1}
\end{figure*}
\begin{figure*}
\centering
\includegraphics[width=0.32\textwidth]{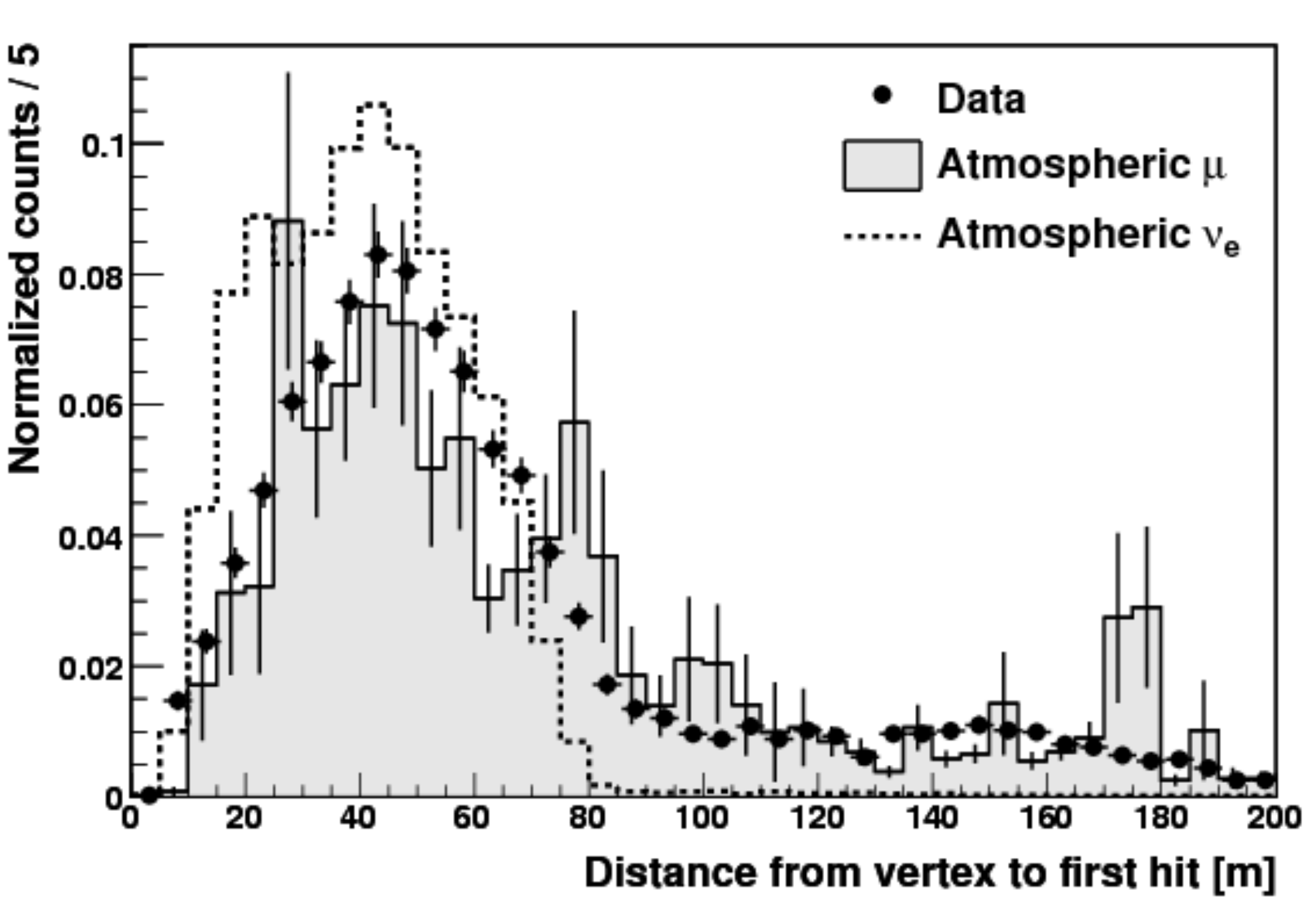}
\includegraphics[width=0.32\textwidth]{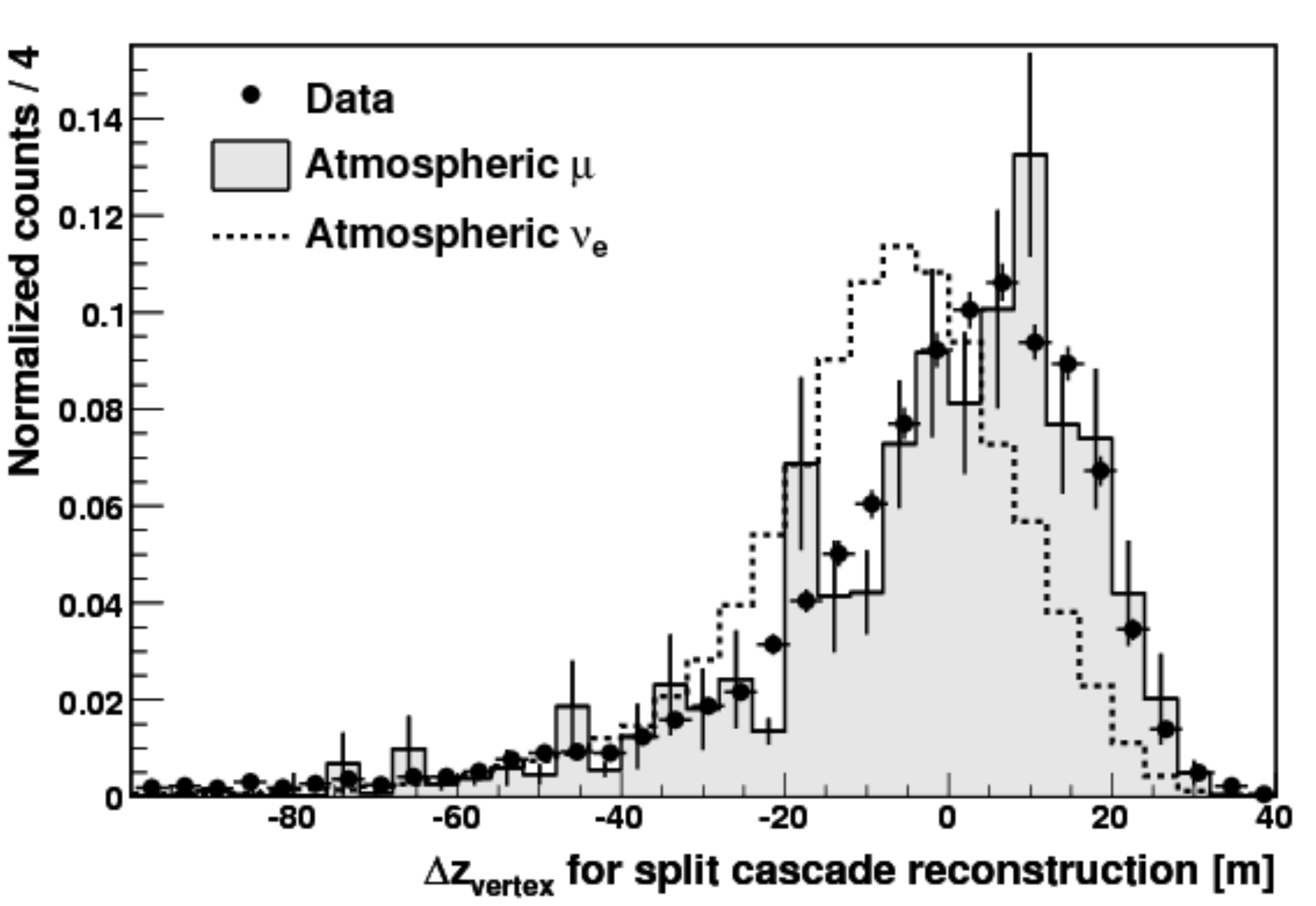}
\includegraphics[width=0.32\textwidth]{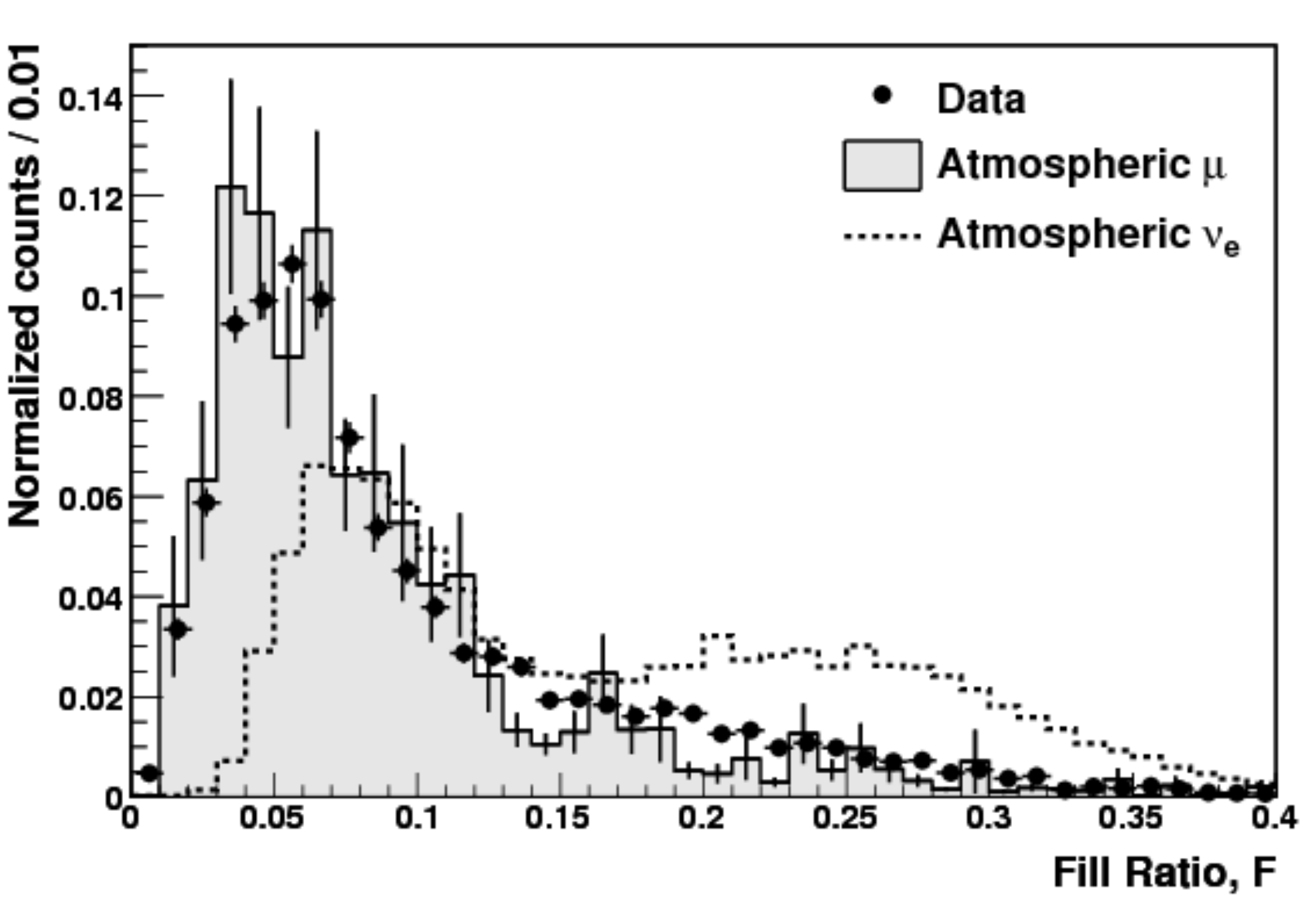}
\caption{Normalized distributions (after Level-4 atmospheric selections) of 
(left panel) distance from the cascade vertex to the first hit in the event (middle panel) difference in $z$ vertex positions for the two split cascade reconstructions
(right panel) fill-ratio from the mean hit distance.}
\label{fig:distlevel4a2}
\end{figure*}

\begin{figure}[h!]
\centering
\includegraphics[width=0.38\textwidth]{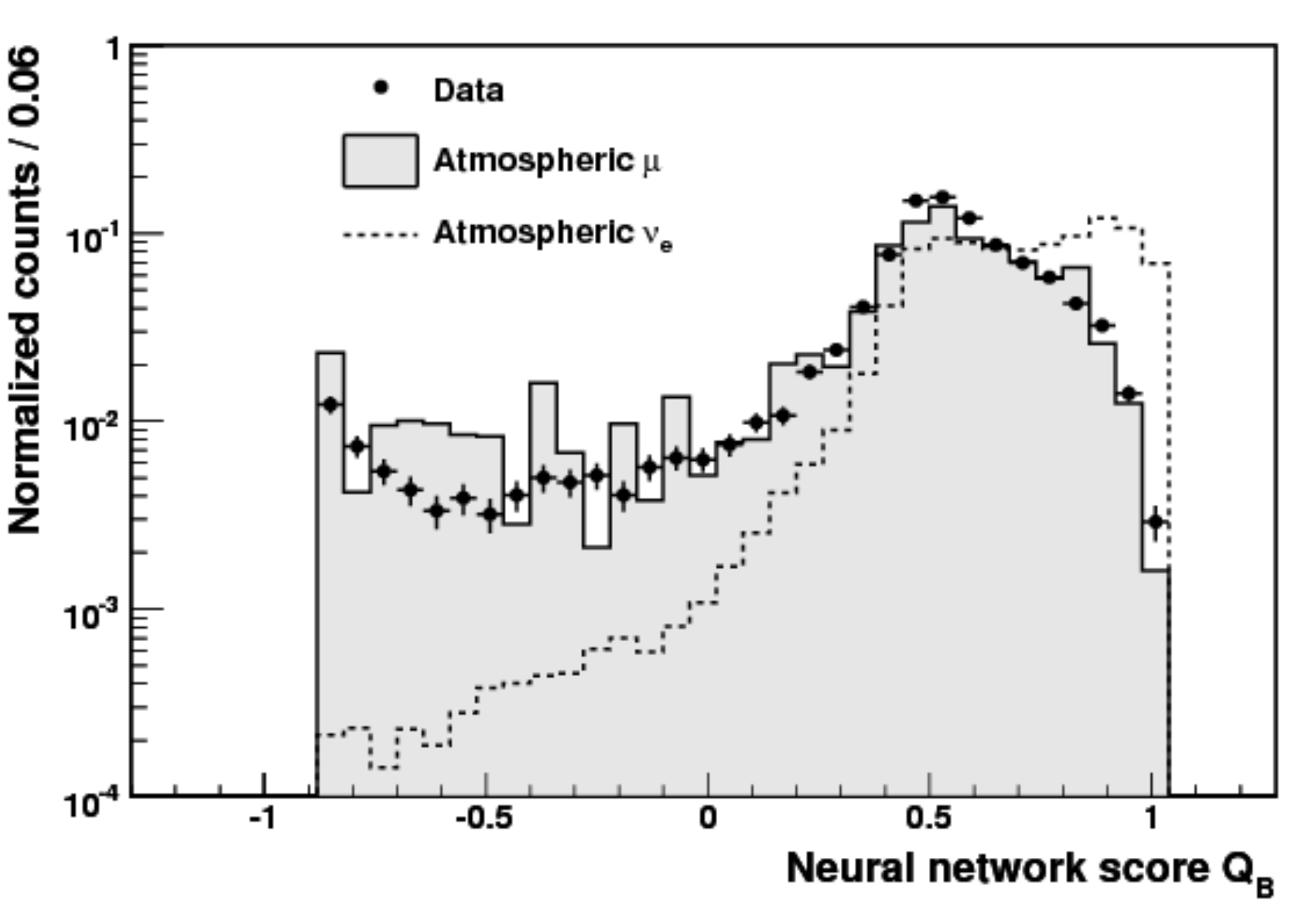}
\includegraphics[width=0.38\textwidth]{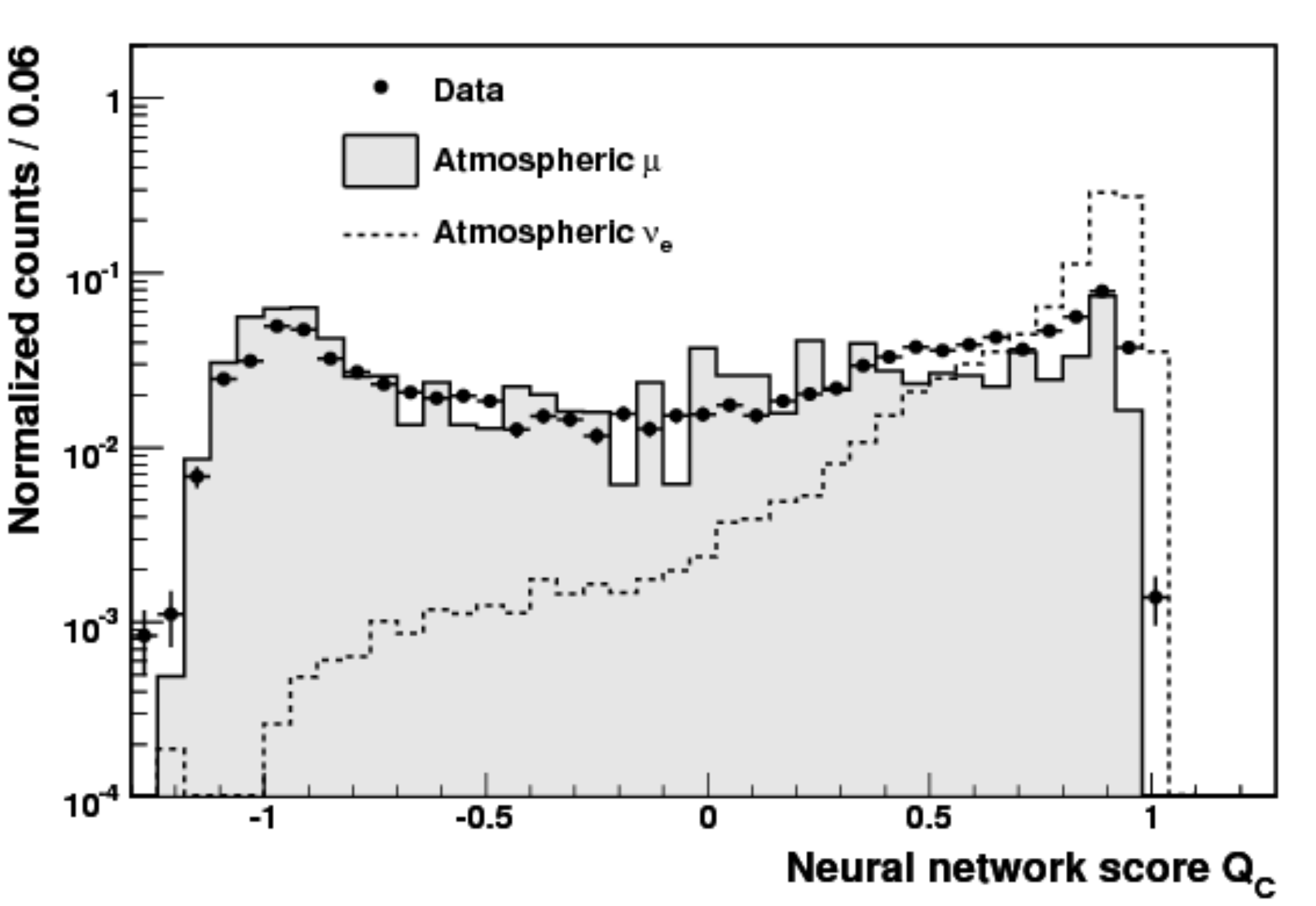}
\caption{Normalized distributions of  
the neural network \\
classification scores (top) ${\rm{Q_B}}$ and (bottom) ${\rm{Q_C}}$.
}
\label{fig:mlplevel4a}
\end{figure}

At Level-5, we chose to train two individual neural networks separately because of limited remaining statistics in the Monte Carlo training samples and to achieve adequate performance.
The first neural network was trained with the number of direct hits from the reconstructed cascade vertex, the total number of observed photoelectrons in all DOMs, 
the difference between log-likelihoods from the cascade vertex reconstruction and from the 32-fold iterative muon track reconstruction, and the reconstructed cascade energy.  
These four variables correlate in a complex way that merits the use of a neural network.
The second neural network was trained with six variables:
the distance from the cascade vertex to the first hit in the event,
the fill-ratio $F$,
the reduced log-likelihood parameter from the cascade vertex reconstruction,
the difference in $z$ vertex positions for the two split cascade reconstructions,
the track zenith angle of the 32-fold iterative muon reconstruction,
and  $\alpha$ defined and used before at Level-4.
Figures~\ref{fig:distlevel4a1} and~\ref{fig:distlevel4a2}
show the distributions of selected input variables, whereas Fig.~\ref{fig:mlplevel4a} shows the two neural network classification scores ${\rm{Q_B}}$ and  ${\rm{Q_C}}$.

The three neural network classification scores, ${\rm{Q_A, Q_B}}$ and ${\rm{Q_C}}$
were optimized for $E_\mathrm{reco} >   5\,\mathrm{TeV}$,  and their product,  
${\rm{Q^{\star} = Q_A \times Q_B \times Q_C}}$,
was used at Level-5. 
In addition, events with a reconstructed cascade vertex in the topmost 60 meters of the detector were rejected.
Multi-muon background  from coincident cosmic ray air showers 
was efficiently rejected by a relatively loose selection criterium on the reduced likelihood parameter from the cascade vertex reconstruction.
The final selection  was
${\rm{Q^{\star}}} > 0.73$ for $E_\mathrm{reco} >   5\,\mathrm{TeV}$.

\subsection{Normalization and Systematic Uncertainties}\label{sec:systematics}

The extraterrestrial and atmospheric analyses have common sources of systematic uncertainty.
The largest contributions to our systematic uncertainty estimate arise from our limited knowledge of the optical properties of the ice and from uncertainties in the cosmic-ray flux and composition.
Other significant contributions result from uncertainty in the DOM detection efficiency and from uncertainties in the neutrino cross-sections and the light output from the cascades.
These sources affect signal and background estimates differently. We describe signal and background separately below.

\begin{figure*}[htb]
\centering
\includegraphics[width=0.335\textwidth]{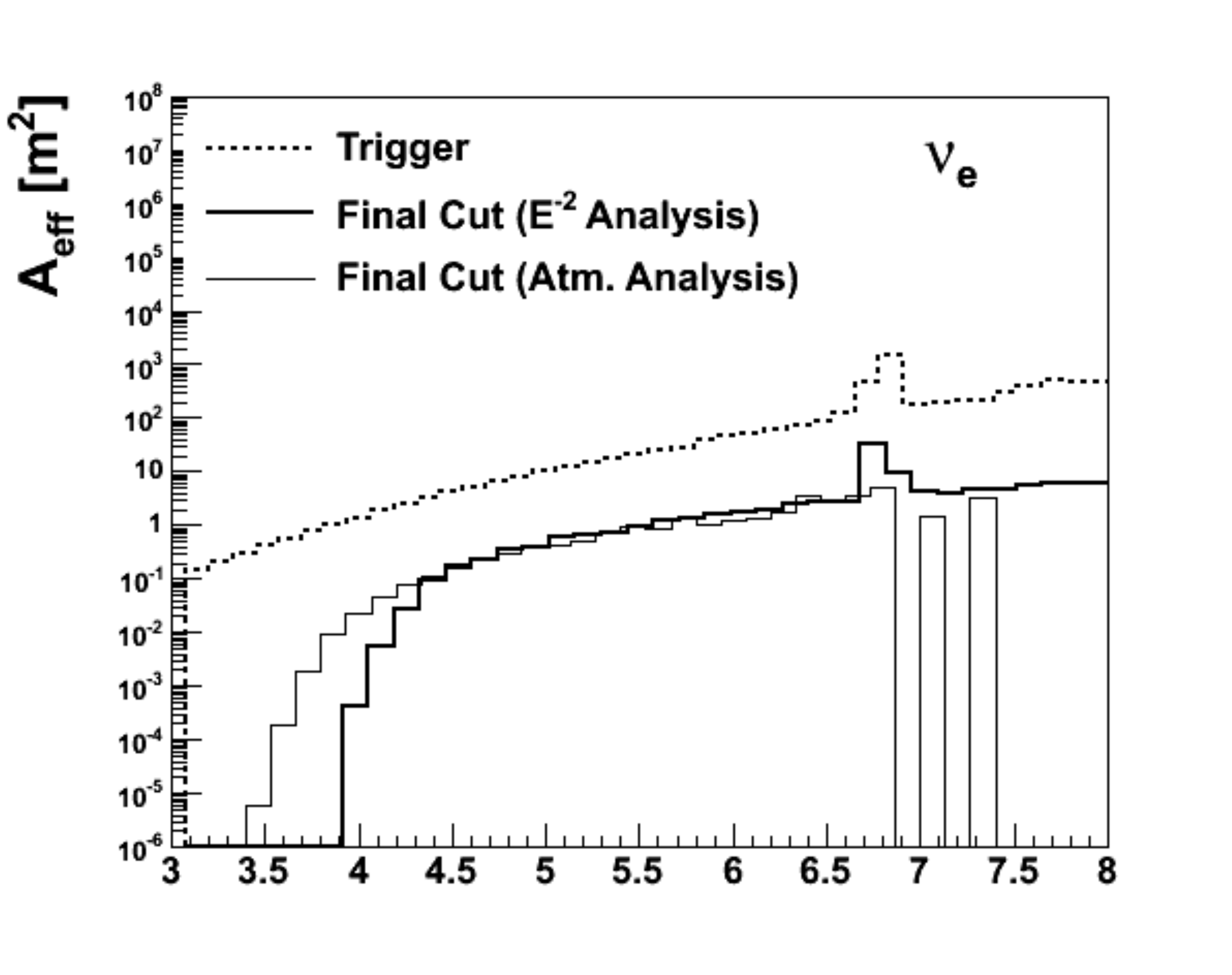}
\hspace*{-0.6cm}\includegraphics[width=0.335\textwidth]{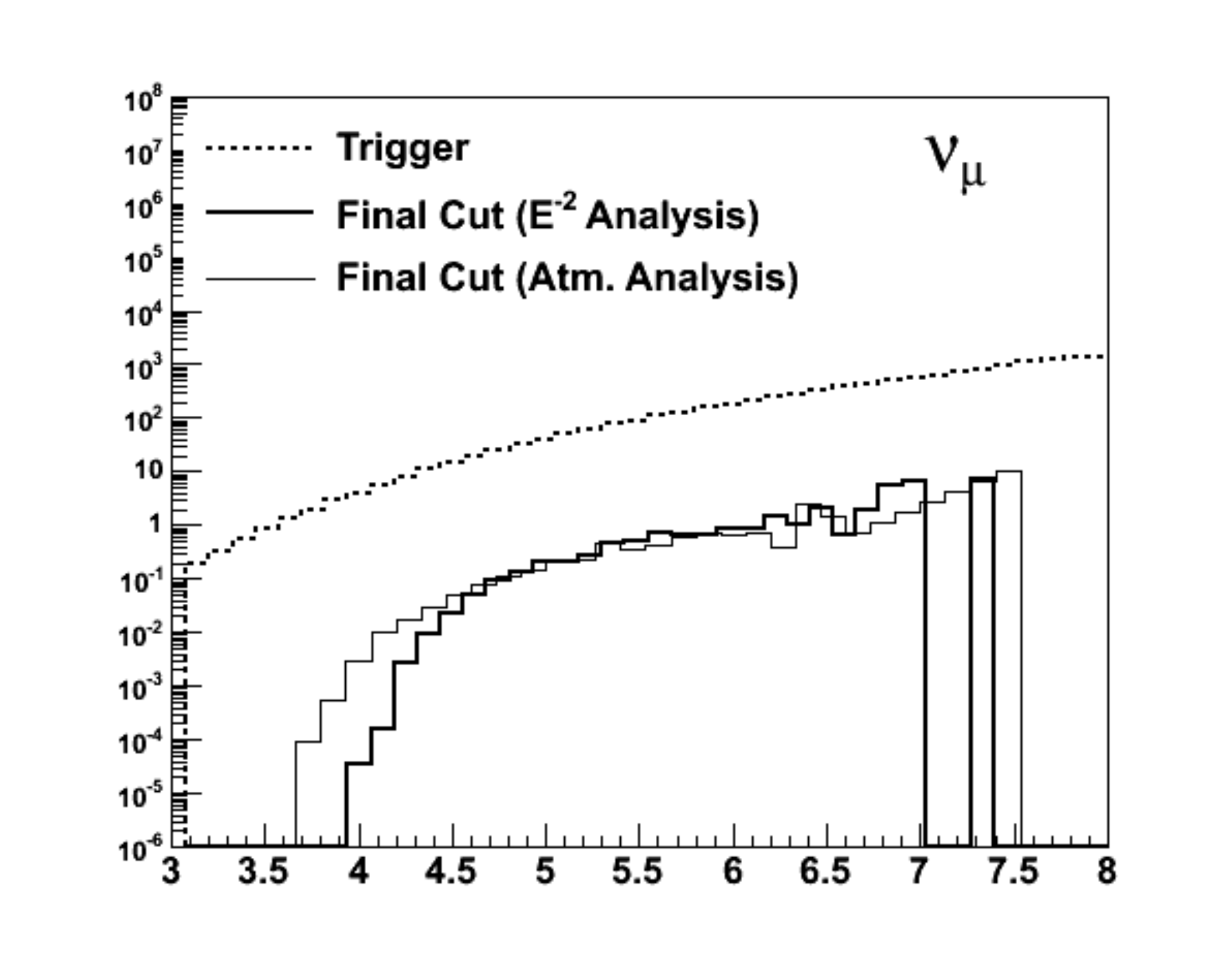}
\hspace*{-0.6cm}\includegraphics[width=0.335\textwidth]{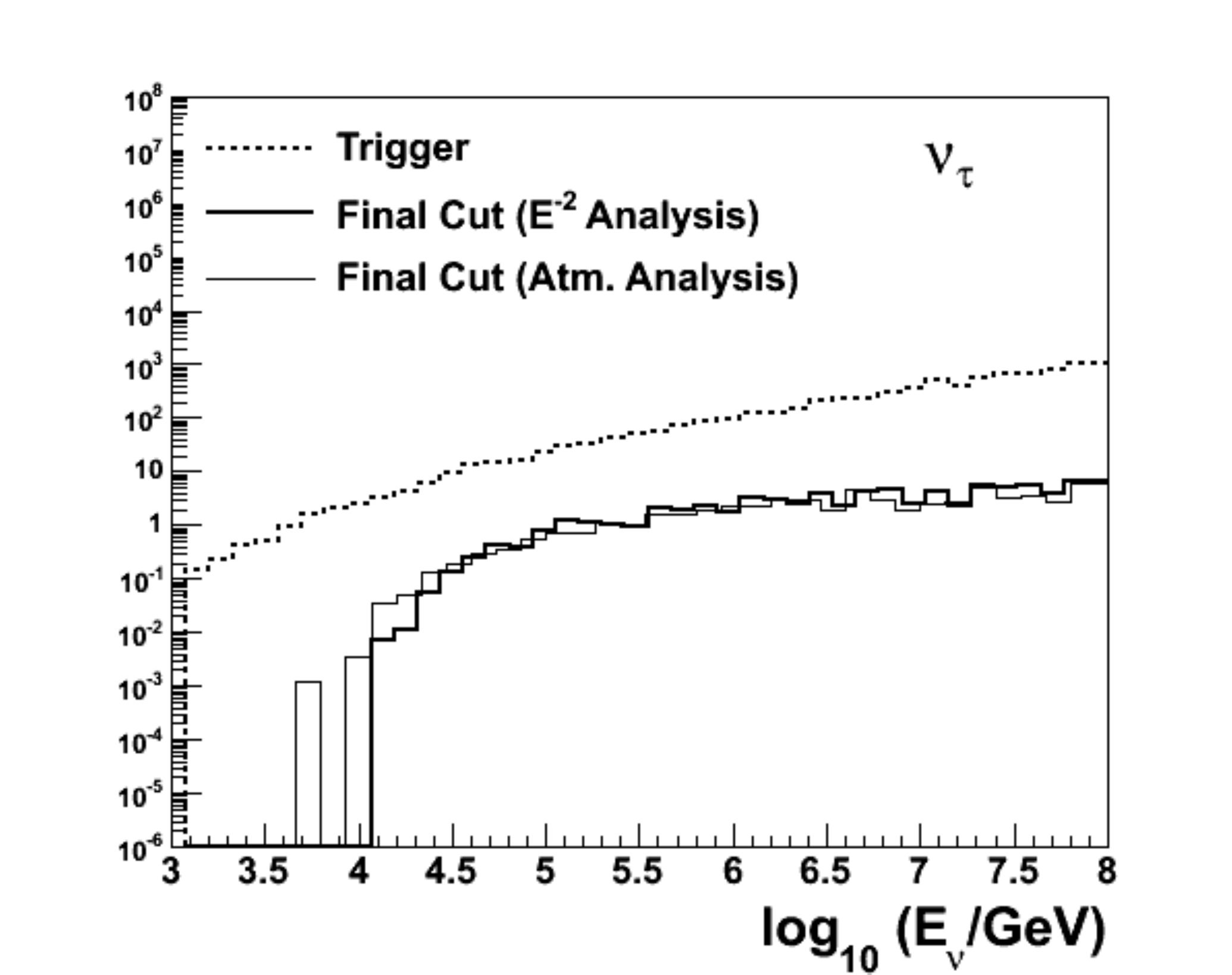}
\caption{The effective area, $A_\mathrm{eff}$, for (left panel) $\nu_e$, (middle panel) $\nu_\mu$, and (right panel) $\nu_\tau$ 
versus neutrino energy at trigger level (dashed curves) and at analysis level after all selections for both the extraterretrial and the atmospheric analysis (continuous curves).  
}
\label{fig:effarea}
\end{figure*}

\subsubsection*{Muon Background Systematics}\label{subsec:muon}
CORSIKA Monte Carlo simulations adequately reproduce the shape of the observed spectra in both analyses but systematically underestimate the absolute rate. 
In the extraterrestrial analysis, an empirical factor to normalize the rate of simulated background to the corresponding data 
was found  to be $3\pm1$,  where the uncertainty covers the difference between the data and CORSIKA  rates  at   selection levels up to  6 (see Table~\ref{tbl:rates}). 
Background events  outnumber signal events by at least 3 orders of magnitude in these event samples.
In the atmospheric analysis, the same factor $3 \pm 1$ was found for 
$\mathrm{Q^{\star}} < 0.4$, where any cascade signal is expected to be vanishingly small.  The factor was the same, within uncertainties, at larger 
$\mathrm{Q^{\star}}$ values.

Simulations based on an alternate ice model~\cite{spice} were performed to assess the possible effects on the empirical normalization factor caused by incomplete 
description of the optical ice properties and necessary approximations in their implementation in the simulations. 
The alternate model is based on a global fit to
recorded charges in very bright events generated by IceCube LED
sources sampling all detector depths.
The background rates simulated with the alternate ice model differed from the rates observed in the data by a factor of two. This is consistent, to within the assigned uncertainties, 
with the difference between the rates simulated using the calibrated ice model and those observed in the data.

The remaining difference in simulated and observed rate is ascribed to
the combined effect of uncertainties in the cosmic ray flux for
protons in the energy range near the knee and in the absolute energy scale.

\subsubsection*{Neutrino Systematics}
 
Simulations show that neutrino events are much less affected by uncertainty in the ice model.  We estimate a $20$\% uncertainty in the rates by comparing simulations with the calibrated and alternate ice models for both atmospheric and $E^{-2}$ energy spectra.

The uncertainty in the DOM sensitivity is taken as $8$\%, based on the measured uncertainty in the PMT sensitivity~\cite{PMTpaper}.
For an $E^{-2}$ neutrino energy spectrum, coupled with a neutrino interaction probability that scales with $E$, this is equivalent to a 8\% uncertainty in flux.
For atmospheric neutrinos, the spectrum can be approximated by $E_\nu^{-3.7}$ for conventional neutrinos and by $E_\nu^{-2.7}$ for ÔpromptÕ neutrinos from 
charm and bottom quark decays. 
These lead to uncertainties of 20\% and 12\%, respectively, for the detector sensitivity.
We conservatively use the larger value for all atmospheric neutrinos.

Uncertainty in the atmospheric neutrino flux forms an additional uncertainty in the 
atmospheric neutrino background to the extraterrestrial cascade search.
The expected rates of atmospheric neutrinos 
are based on predictions from Ref.~\cite{barr} and Ref.~\cite{naumov} for the conventional and prompt components, respectively. 
We assign a 20\% uncertainty in these rates~\cite{OxanaPaper}.  

In the relevant energy range, uncertainty in the neutrino interaction cross-sections is about 6\%, caused largely by uncertainties in the parton densities \cite{AMANDAIIpointsource,IC22EHEpaper}.  

The total uncertainty in the atmospheric (extraterrestrial) neutrino detection efficiency was estimated to be 29\% (22\%) by adding these contributions in quadrature, see Table~\ref{Tab:systematics}.

 \begin{table}[h]
	\centering
	\caption{Table of systematic uncertainties for the atmospheric and extraterrestrial neutrinos.}
	\label{Tab:systematics}
\begin{tabular}{cccc}
	\hline\hline
Source                                                       & $\nu^{\rm{atm}}$   & $\nu({E^{-2}})$ \\   
        \hline
Ice Properties                                                        &  $20$\%                          & $20$\%              \\
DOM Efficiency                                                                   &  $20$\%                     & $8$\%              \\
Neutrino Cross-Section                                           &  $6$\%                     & $6$\%                 \\
	\hline 
Total Uncertainty                                                   &   $29$\%                        & $22$\% \\
	\hline\hline
\end{tabular}
\end{table}

To check the energy scale, we studied the detector response to a 337 nm $N_2$ laser, known as the Ôstandard candle.Õ  This laser is on one of the IceCube strings, at a depth of 2153 m.  It produces light that is shaped like a Cherenkov cone pointing downward; the light output is calibrated to $\pm 10\%$.  For DOMs that are far enough from the laser to avoid saturation (defined as observing less than 20,000 photoelectrons), the total observed charge is 3\% lower than our Monte Carlo prediction for the calibrated ice model, and 12\% lower than the expectation for the alternate ice model.   This is well within the expectations from the ice model and the DOM efficiency uncertainty.  

\section{Results}\label{sec:results}
After the event selections had been finalized, the analyses were unblinded. The 10\% of the data, which was used to develop the selections, 
was discarded from the results.  The remaining data correspond to 257 days of livetime.

The effective area $A_{\rm{eff}}$, defined as the equivalent area with 100\% neutrino interaction probability and detection efficiency, was obtained by passing simulated signal Monte Carlo events through the analyses.
Figure \ref{fig:effarea} shows $A_\mathrm{eff}$ for $\nu_e$, $\nu_\mu$, and $\nu_\tau$ versus neutrino energy.
The contribution from the Glashow resonance is clearly visible in $A_\mathrm{eff}$ for $\nu_e$ at trigger level and at final selection level for the extraterrestrial analysis. 
The analyses cover complementary energy ranges, with the atmospheric analysis naturally covering lower energies than the extraterrestrial analysis.  The analyses have similar $A_\mathrm{eff}$ in the region of overlap. 

The energy resolution is $\Delta(\log_{10}E_{\nu}) \sim0.26\,(0.18)$  for the $E^{-2}$ (atmospheric) $\nu_{e}$ energy spectrum.  The $x$, $y$ and $z$ position resolution is $\sim10$ meters.  

\subsection{Atmospheric Neutrino Results}
 
Before unblinding, the event selection criteria for the atmospheric analysis were optimized for a conventional atmospheric neutrino flux 
with a 5 TeV threshold on the reconstructed cascade energy. 
Twelve events remained in the 90\% of the recorded data 
after an a-posteriori  selection criterium on the multivariate selection variable 
$\mathrm{Q^{\star}}$  at 0.9.  The expected atmospheric electron and muon neutrino signal is  $7.0\pm2.0$ events for these selection criteria. 
The prompt neutrino component is $25\%$ of the expectation.  
Only one standard CORSIKA Monte Carlo event passed the final selections.  
Muon background estimate based on this one Monte Carlo event  is  $9\pm9$. 
It  includes  an empirical scaling factor 
to normalize the simulated background to the data. 
The limited  background Monte Carlo statistics preclude detection of a signal.

\begin{figure}[h]
\centering
\includegraphics[width=0.4\textwidth]{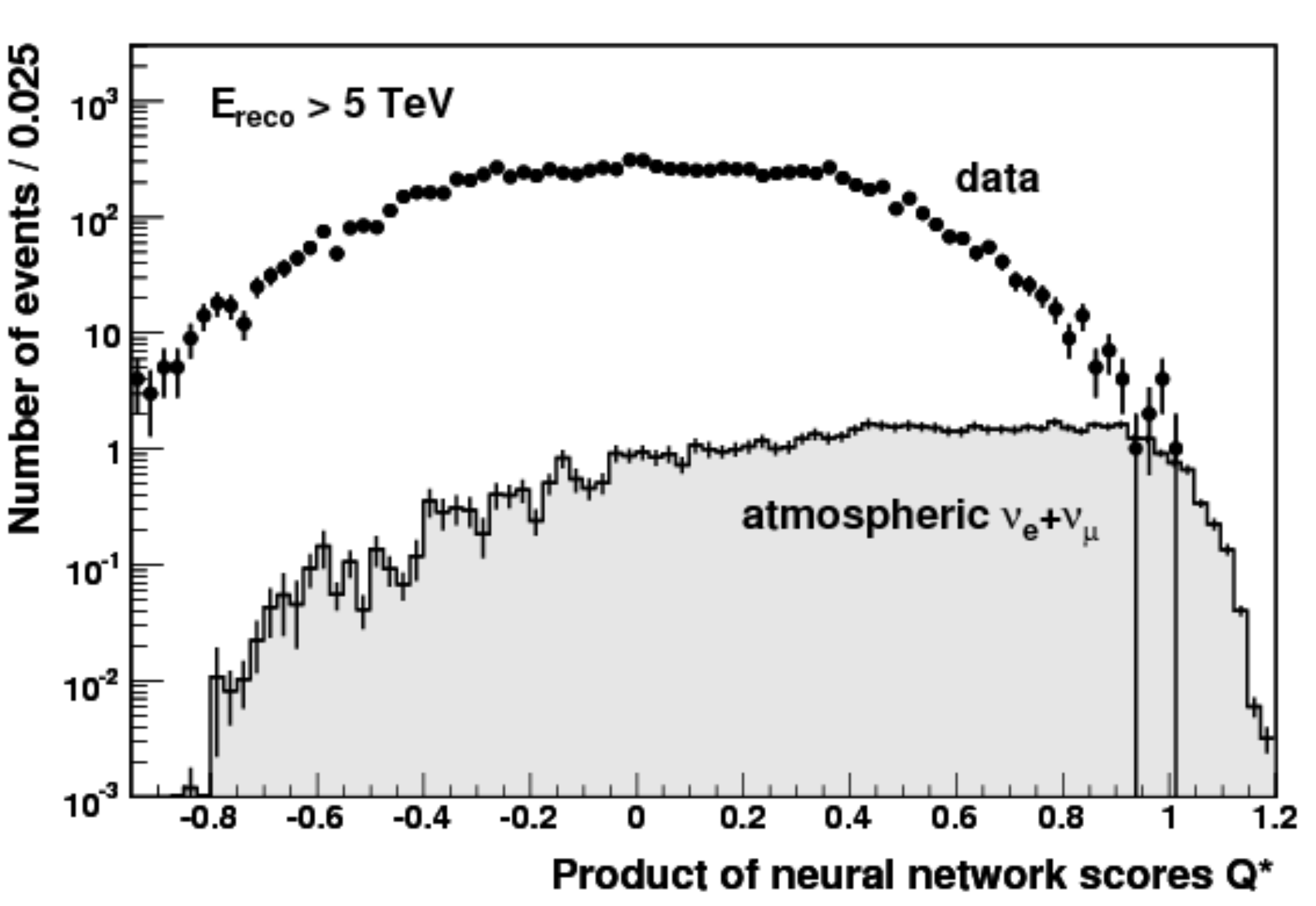}
\caption{Distribution of the multivariate selection parameter 
$\mathrm{Q^{\star}}$  for data and for a simulated cascade signal from $\nu_e$ and $\nu_\mu$ with reconstructed energies above 5 TeV in the atmospheric analysis.}
\label{atmnu-final-dist}
\end{figure}

Figure~\ref{atmnu-final-dist} shows the distribution of the variable $\mathrm{Q^{\star}}$ for the full data set with $E_\mathrm{reco}>5$~TeV and compares it to the corresponding distribution for a
simulated signal of cascades from atmospheric  electron and muon neutrinos.

\begin{figure}[h]
\centering
\includegraphics[width=0.4\textwidth]{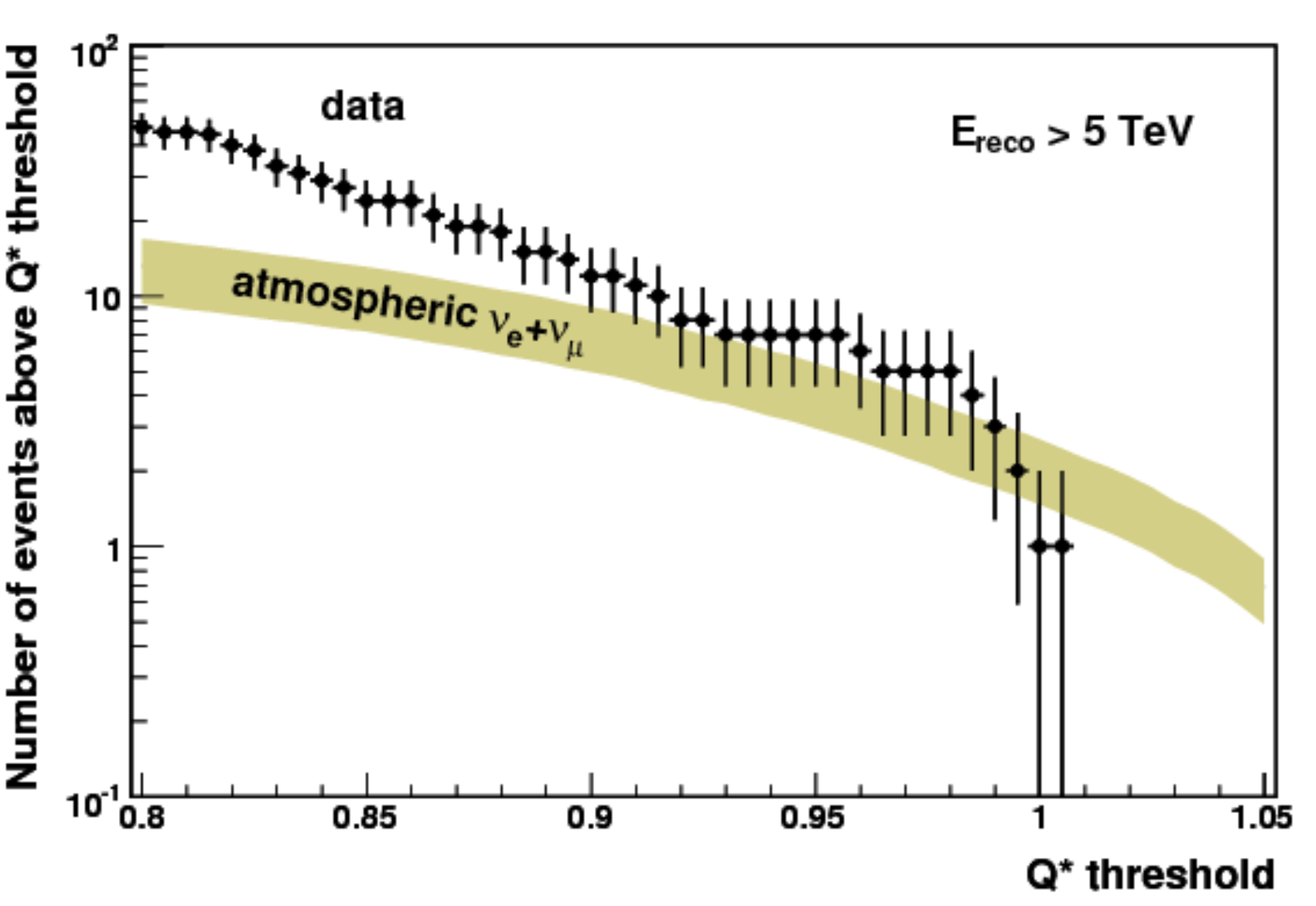}
\includegraphics[width=0.4\textwidth]{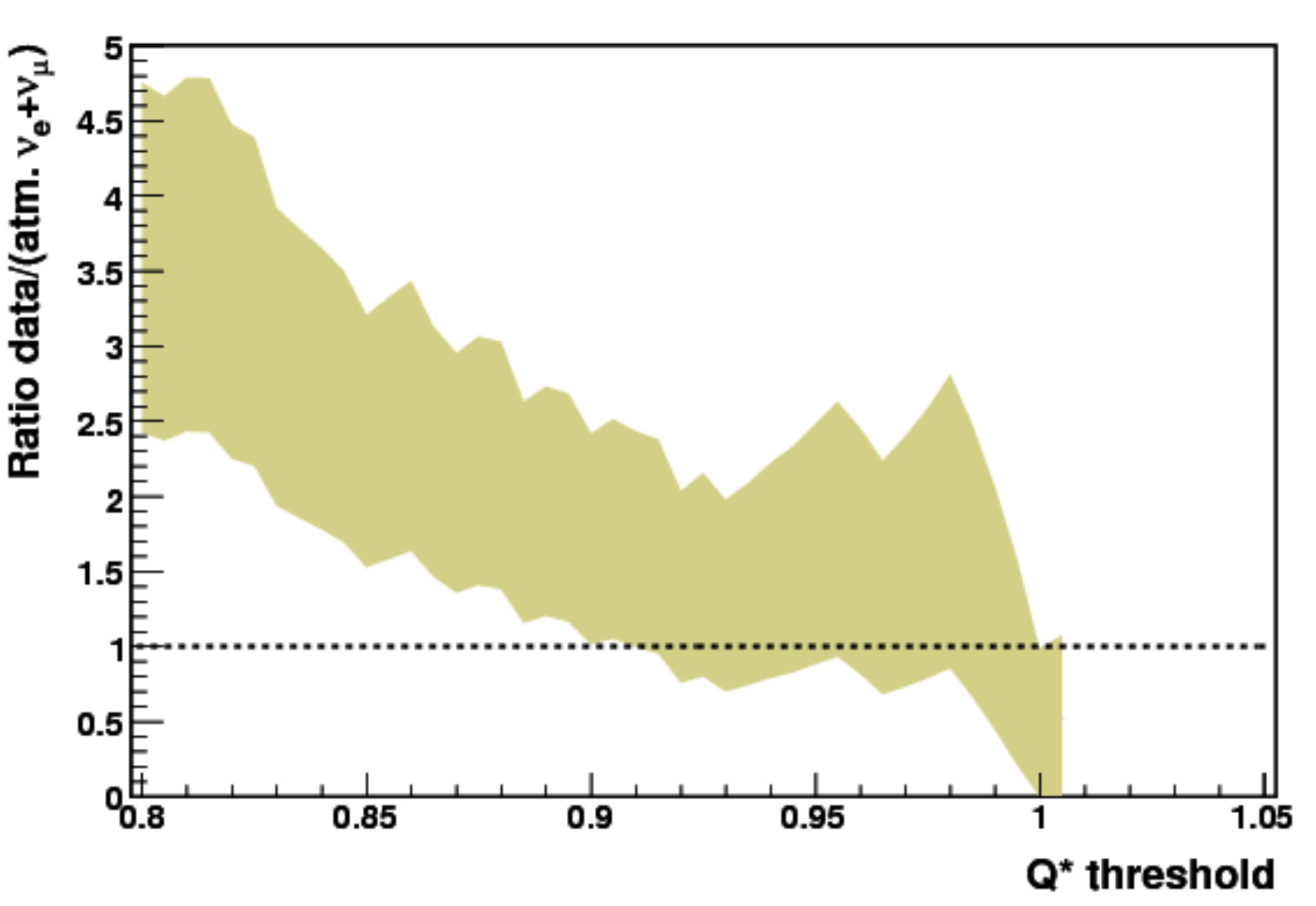}
\caption{Experimental events reconstructed with energies above 5~TeV in the atmospheric analysis compared to an expected cascade signal from atmospheric neutrinos. 
(top) The cumulative number of events above a threshold in the  multivariate selection parameter 
$\mathrm{Q^{\star}}$ for data and simulated signal.  (bottom) The ratio of data to expected signal as a function of selection threshold. 
}
\label{atmnu-final-cumu}
\end{figure}

Figure~\ref{atmnu-final-cumu} shows the cumulative number of events as a function of the threshold applied on the 
$\mathrm{Q^{\star}}$ variable. The top panel compares the experimental data to the
signal expectation from simulations, shown as a band with a width determined by the systematic uncertainty in the atmospheric neutrino flux.
The bottom panel in Fig.~\ref{atmnu-final-cumu} shows the ratio of data events over expected signal events as a function of selection threshold. This ratio crosses unity for the most discriminating
selection criterium that can be applied without rejecting all the data, and where the remaining few events have the highest probabilities of being due to atmospheric neutrino cascades.
This implies that, in using this analysis, the detector becomes sensitive to atmospheric cascade events due to neutrinos, but the exposure is not sufficient for a statistically significant detection.

At a 15 TeV energy threshold and an a-posteriori  selection criterium on the multivariate selection variable 
$\mathrm{Q^{\star}}  > 0.9$, six events remained in the 90\% of the recorded data.  
The expected background from atmospheric muons was estimated to be 
$5.0\pm3.8$ events. This estimate was made with high-energy CORSIKA simulations 
and includes the empirical scaling factor 
to normalize the simulated background to the data.   
The expected atmospheric electron and muon neutrino signal is $3.0\pm0.9$ events for these selection criteria.
The prompt neutrino component is $40\%$ of the expectation.  
The observed number of events is compatible with Monte Carlo prediction. 

\subsection{Diffuse Flux Limit}
A total of $14$ events passed all selection levels in the extraterrestrial analysis with an expected total background of $8.3 \pm 3.6$ events.
The background was estimated from simulations, which indicate sizable contributions from atmospheric muons and from atmospheric neutrinos (see Table~\ref{tbl:rates}).

\begin{figure}
\centering
\includegraphics[height=0.35\textwidth]{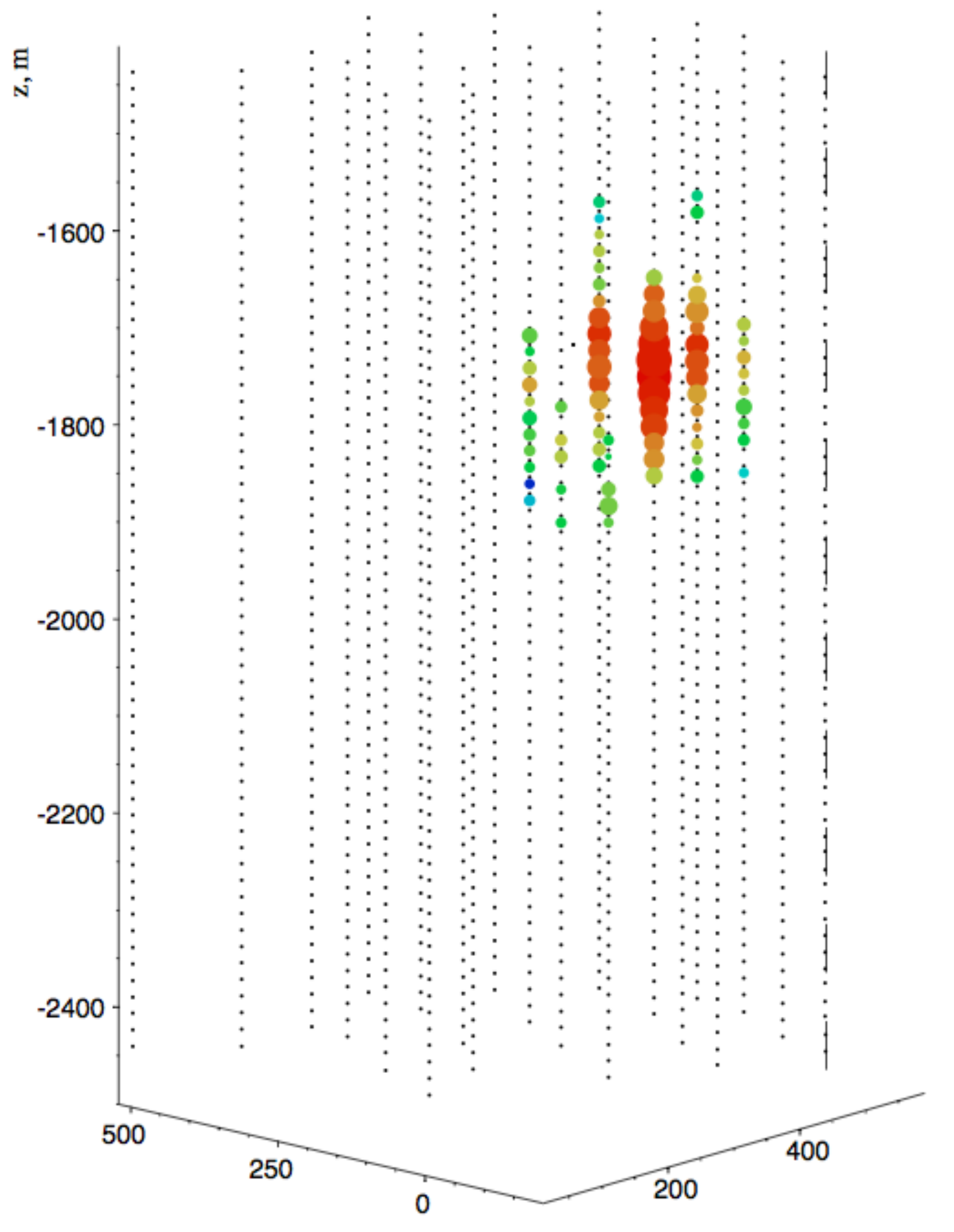}
\caption{(Color online) The MC muon background  event which passed all selections in  extraterrestrial analyses and which has the reconstructed energy of $18~{\mathrm{TeV}}$ is displayed from the side. 
Different colors of the circles represent different  DOMs hit times, with early hits in red and late hits in blue. The size of the circles represents the amplitude.}
\label{event:mc}
\end{figure}

Four events from the high-energy CORSIKA Monte Carlo background sample satisfy all event selection criteria.
They all originated from proton induced showers, with zenith angles of around 70 degrees and energies in the range of $0.5-3.5$ PeV.
Figure~\ref{event:mc} shows  one out of these four Monte Carlo muon background events. 
Taking into account the event weights in the simulation and the rate normalization, the four simulated events that satisfied all selections correspond to $5.4 \pm 3.5$ muon background events in 257 days of livetime.
The number of background atmospheric neutrinos from conventional and prompt sources was estimated to be $2.9 \pm 0.9$,
where the central value $2.9$ was obtained assuming  atmospheric neutrino fluxes from Refs.~\cite{barr,naumov}.

The expected rate of signal neutrino events was evaluated for an assumed flux of
 	$E^{2} \Phi_\mathrm{model} = 1.0 \times 10^{-6} {\rm{GeV \cdot cm^{-2} \cdot s^{-1} \cdot sr^{-1}}}$ 
	for the sum of all flavors and  
using the effective areas, $A_\mathrm{eff}$, given in Figure~\ref{fig:effarea}.
The results are given in Table~\ref{tbl:rates} for each flavor.
In 257 days of livetime, a total of $46 \pm 10$
signal events is expected, assuming that  this $\Phi_{model}$  receives equal contributions from all flavors ($\nu_e : \nu_\mu : \nu_\tau$ flux is $1 : 1 : 1$  at the Earth).
Electron neutrinos  contribute about 
$40\%$, tau neutrinos about $45\%$ and muon neutrinos the remaining $15\%$.

\begin{figure}[h]
\includegraphics[height=0.31\textwidth]{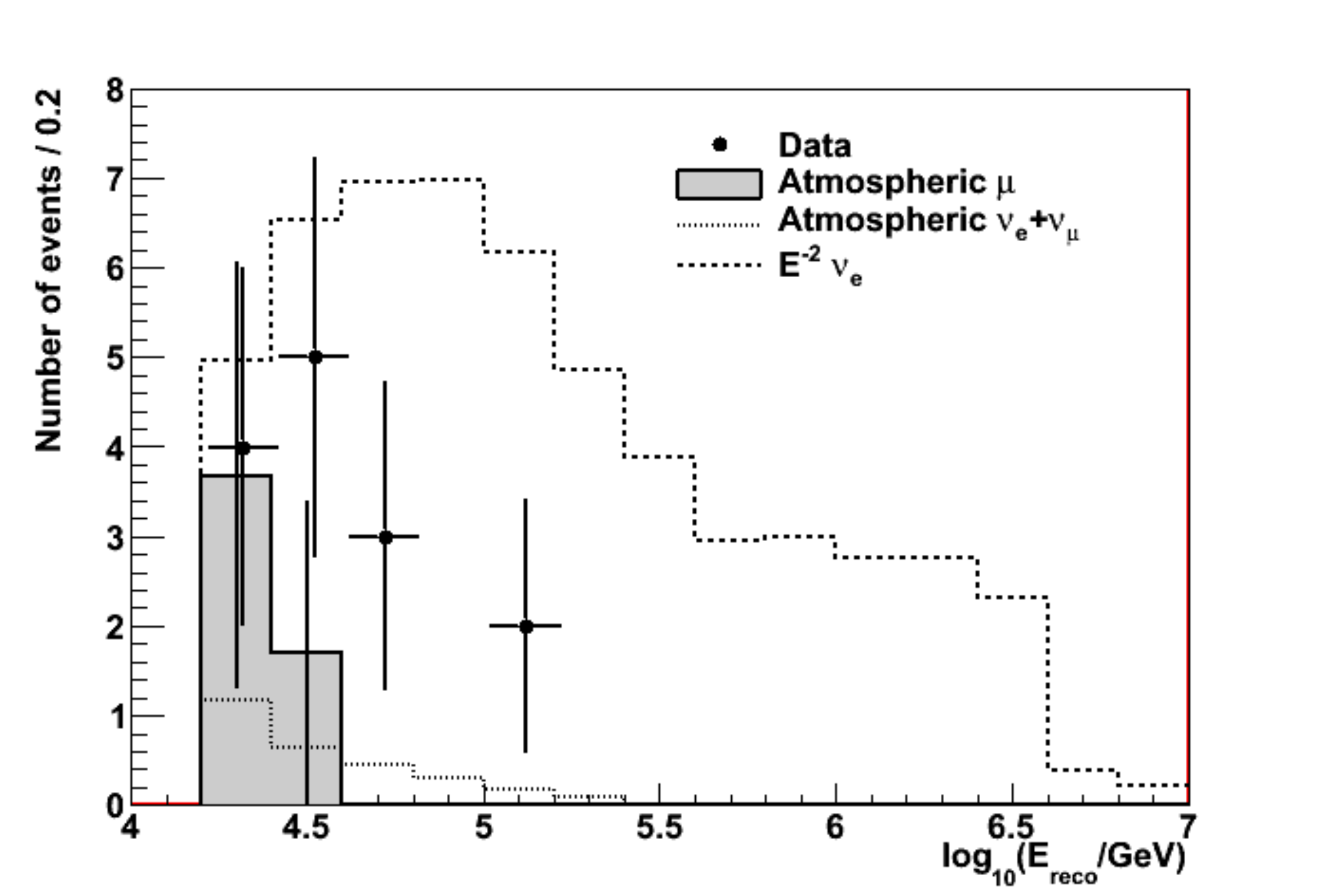}
\caption{Cascade reconstructed energy distribution after all selections in the extraterrestrial analysis. }
\label{fig:finalenergy}
\end{figure}

The energy distributions are shown in  Fig.~\ref{fig:finalenergy}.

By using the Feldman-Cousins method~\cite{feldman-cousins} as implemented in the software package POLE++~\cite{pole++} 
to account for statistical and systematic uncertainties in the estimated background and signal counts, we set an upper limit on the number of signal events at 90\% confidence level of $\mu_{90\%}=16.6$.
This $\mu_{90\%}$ is below the expected total number of signal events  and, hence, $\Phi_\mathrm{model}$ is rejected at 90\% confidence.
Conversely, an upper limit at 90\% confidence can be set at  $E^2 \Phi_{90\%~\mathrm{C.L.}} < 3.6\times10^{-7}\,\mathrm{GeV \cdot cm^{-2} \cdot s^{-1} \cdot sr^{-1}}$ 
on the diffuse flux of neutrinos of all flavors assuming that $\Phi \propto E^{-2}$ and that the flux at the detector receives equal contributions from all flavors.

An upper limit on the flux of $\nu_e$ that does not depend on the assumption of equal flux contributions for each flavor was derived by assuming that the $\nu_\mu$ and $\nu_\tau$ fluxes are zero. 
This upper limit on the flux of electron neutrinos is
 $E^2 \Phi_{90\%~\mathrm{C.L.}} < 3.0\times10^{-7}\,\mathrm{GeV \cdot cm^{-2} \cdot s^{-1} \cdot sr^{-1}}$.

\begin{figure}[h]
\includegraphics[height=0.5\textwidth,angle=90]{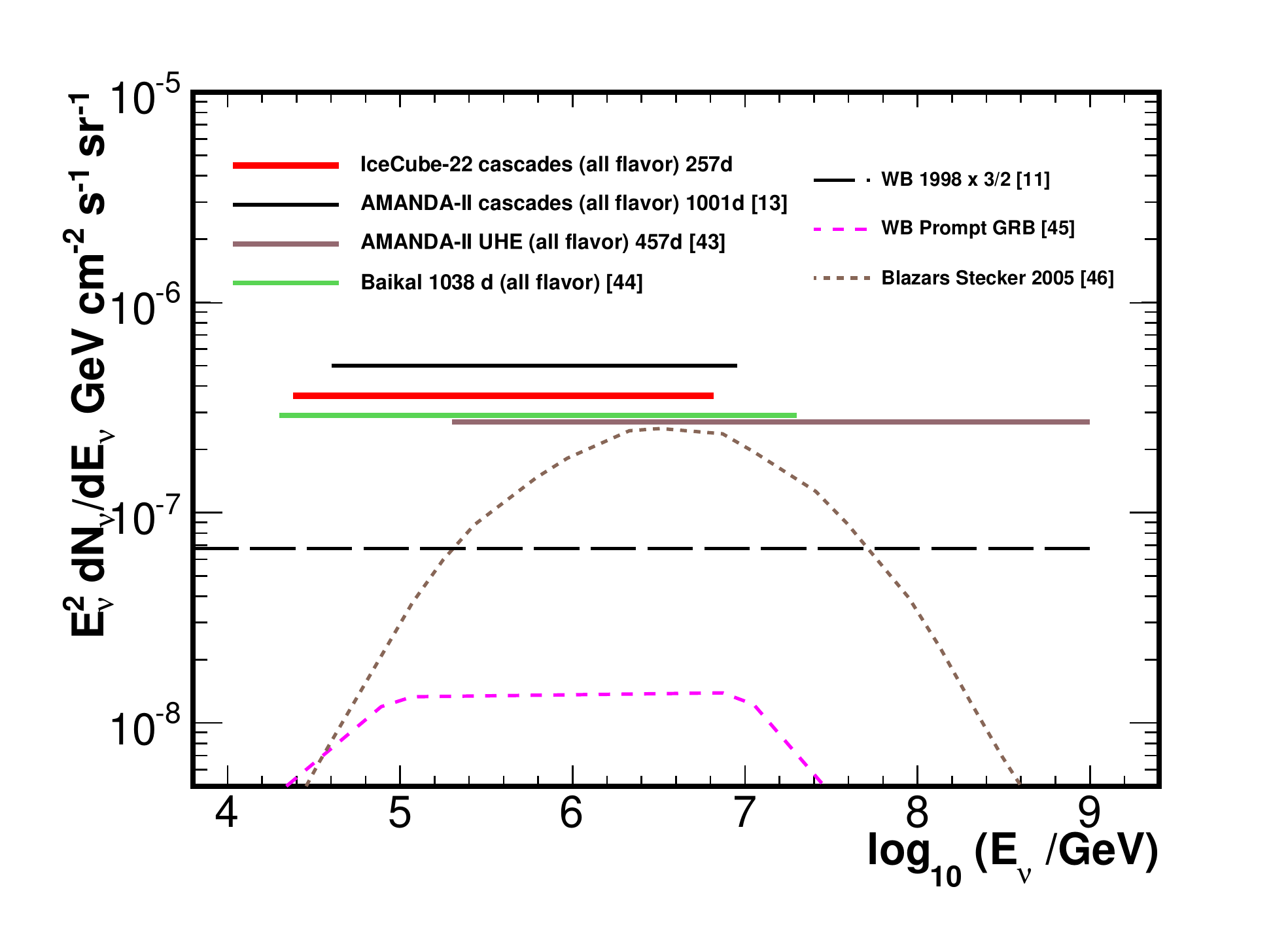}
\caption{  
(Color online) Experimental upper limits on the diffuse flux of neutrinos from sources with $\Phi \propto E^{-2}$  energy spectrum
and theoretical predictions for neutrino fluxes from astrophysical neutrino sources. 
}
\label{fig:limit}
\end{figure}

In these limits, the central $90\%$ of $\nu$ signal events are in the energy range from $24$\,TeV to $6.6$\,PeV with a mean energy of $\sim220$\,TeV. 
The limits are shown in Fig.~\ref{fig:limit}. \newline

\begin{figure}
\centering
\includegraphics[height=0.35\textwidth]{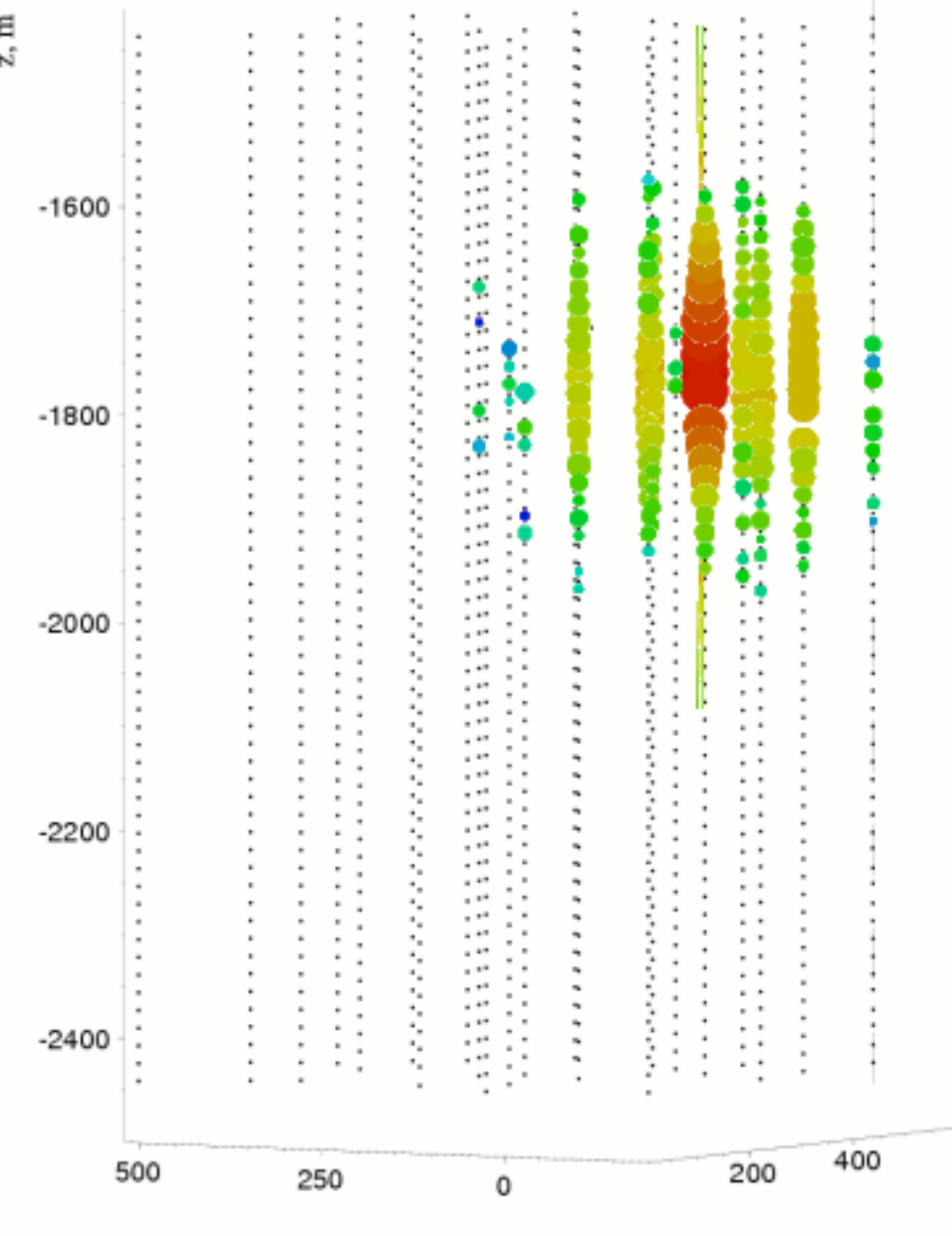}
\caption{(Color online) The data event which passed all selections in atmospheric and extraterrestrial analyses and which has the highest reconstructed energy of $134~{\mathrm{TeV}}$ is displayed from the side.  
Different colors of the circles represent different  DOMs hit times, with early hits in red and late hits in blue. The size of the circles represents the amplitude.}
\label{fig:eventdisplay}
\end{figure}

Figure~\ref{fig:eventdisplay} shows the experimental event with the highest reconstructed energy that passed all selections in both atmospheric and extraterrestrial analyses.
 
\section{Summary}\label{sec:summary}
In summary, we report the first search for cascades induced by atmospheric and by diffuse astrophysical neutrinos with the IceCube detector.
The data, obtained in 2007--2008 with a configuration of 22 active strings, amount to  
257 days of livetime and   
was searched for charged current interactions of $\nu_e$ and $\nu_\tau$, and for  neutral current interactions of neutrinos of all flavors.

The atmospheric neutrino analysis used neural-network based event selections and resulted in a total of $12$ candidate events with energies above $5$ TeV after event selections.
Within the large uncertainties, the observed number of events is consistent with the expected background.
 
The astrophysical neutrino analysis used one and two dimensional selection criteria  and was optimized for higher energies than the atmospheric neutrino analysis.
A total of 14 events with energies above 16\,TeV remained after event selections, with an expected total background contribution of $8.3 \pm 3.6$ events.
We derive an upper limit at 90\% confidence of $E^2 \Phi_{90\%CL} < 3.6\times10^{-7}\,\mathrm{GeV \cdot cm^{-2} \cdot s^{-1} \cdot sr^{-1}}$ on the diffuse flux of 
astrophysical neutrinos with the assumption that the energy spectrum $\Phi \propto E^{-2}$ and that the flavor composition of the $\nu_e : \nu_\mu : \nu_\tau$ flux is $1 : 1 : 1$ at the Earth.
In this limit, 90\% of the expected signal events have energies between $24$\,TeV and $6.6$\,PeV.

This is below the limit that was recently reported from final AMANDA data, corresponding to 1001 days of livetime~\cite{OxanaPaper}.
Once construction is completed in 2011, IceCube will consist of 86 strings covering a volume of $1\,\mathrm{km}^3$.
Future IceCube searches will thus benefit from a considerably larger size and are expected to have significantly improved detection sensitivity.

\begin{acknowledgments}

We acknowledge the support from the following agencies:
U.S. National Science Foundation-Office of Polar Programs,
U.S. National Science Foundation-Physics Division,
University of Wisconsin Alumni Research Foundation,
the Grid Laboratory Of Wisconsin (GLOW) grid infrastructure at the University of Wisconsin - Madison, the Open Science Grid (OSG) grid infrastructure;
U.S. Department of Energy, and National Energy Research Scientific Computing Center,
the Louisiana Optical Network Initiative (LONI) grid computing resources;
National Science and Engineering Research Council of Canada;
Swedish Research Council,
Swedish Polar Research Secretariat,
Swedish National Infrastructure for Computing (SNIC),
and Knut and Alice Wallenberg Foundation, Sweden;
German Ministry for Education and Research (BMBF),
Deutsche Forschungsgemeinschaft (DFG),
Research Department of Plasmas with Complex Interactions (Bochum), Germany;
Fund for Scientific Research (FNRS-FWO),
FWO Odysseus programme,
Flanders Institute to encourage scientific and technological research in industry (IWT),
Belgian Federal Science Policy Office (Belspo);
University of Oxford, United Kingdom;
Marsden Fund, New Zealand;
Japan Society for Promotion of Science (JSPS);
the Swiss National Science Foundation (SNSF), Switzerland;
A.~Gro{\ss} acknowledges support by the EU Marie Curie OIF Program;
J.~P.~Rodrigues acknowledges support by the Capes Foundation, Ministry of Education of Brazil.

\end{acknowledgments}

\newpage 

\bibliographystyle{apsrev}

\end{document}